\documentclass[aps,prx,twocolumn,superscriptaddress,nofootinbib]{revtex4-2}
\usepackage{amsmath,amssymb,amsfonts}
\usepackage{graphicx}
\usepackage{hyperref}
\usepackage{xcolor}
\usepackage{placeins}

\usepackage{amsthm}

\usepackage{fancyhdr}
\usepackage[calcwidth]{titlesec}
\usepackage[font={footnotesize},labelfont={bf}]{caption}
\usepackage{setspace}

\usepackage{extarrows}

\usepackage[most]{tcolorbox}
\newtcolorbox{myt}[2][]{%
  attach boxed title to top center
               = {yshift=-4pt},
  colback      = blue!5!white,
  colframe     = blue!75!black,
  halign       = flush left,
  fonttitle    = \bfseries\sffamily,
  colbacktitle = blue!65!black,
  title        = #2,#1,
  enhanced,
}
\newtcolorbox{myd}[2][]{%
  attach boxed title to top center
               = {yshift=-4pt},
  colback      = violet!5!white,
  colframe     = violet!75!black,
  halign       = flush left,
  fonttitle    = \bfseries\sffamily,
  colbacktitle = violet!65!black,
  title        = #2,#1,
  enhanced,
}
\newtcolorbox{mye}[2][]{%
  attach boxed title to top center
               = {yshift=-4pt},
  colback      = purple!5!white,
  colframe     = purple!75!black,
  halign       = flush left,
  fonttitle    = \bfseries\sffamily,
  colbacktitle = purple!65!black,
  title        = #2,#1,
  enhanced,
}

\newtcolorbox{myg}[2][]{%
  attach boxed title to top center
               = {yshift=-4pt},
  colback      = green!5!white,
  colframe     = green!75!black,
  halign       = flush left,
  fonttitle    = \bfseries\sffamily,
  colbacktitle = green!65!black,
  title        = #2,#1,
  enhanced,
}

\providecommand{\U}[1]{\protect\rule{.1in}{.1in}}

\usepackage{dsfont}

\usepackage{mathtools}

\def\distill{{\rm Distill}}
\def\cost{{\rm Cost}}

\def\1{\mathbf{1}}

\def\prob{{\rm Prob}}

\def\spec{{\rm spec}}

\def\pure{{\rm PURE}}
\def\pos{{\rm Pos}}

\def\cptp{{\rm CPTP}}

\def\bdelta{{\boldsymbol{\delta}}}

\newcommand{\epm}{\end{pmatrix}}
\newcommand{\bpm}{\begin{pmatrix}}

\renewcommand{\log}{{\operatorname{log}}}

\newcommand{\ebm}{\end{bmatrix}}
\newcommand{\bbm}{\begin{bmatrix}}

\def\bmyd{\begin{myd}{}
\begin{definition}}
\def\emyd{\end{definition}\end{myd}}

\def\bmyl{\begin{myg}{}
\begin{lemma}}
\def\emyl{\end{lemma}\end{myg}}

\def\bmyt{\begin{myt}{}
\begin{theorem}}
\def\emyt{\end{theorem}\end{myt}}

\def\bmyc{\begin{myg}{}
\begin{corollary}}
\def\emyc{\end{corollary}\end{myg}}

\def\>{\rangle}
\def\<{\langle}

\def\id{\mathsf{id}}

\def\mE{\mathcal{E}}

\def\mF{\mathcal{F}}
\def\mN{\mathcal{N}}

\def\mP{\mathcal{P}}

\def\mV{\mathcal{V}}

\newcommand{\supp}{\operatorname{supp}}

\renewcommand{\geq}{\geqslant}
\renewcommand{\leq}{\leqslant}

\newcommand{\ben}{\begin{enumerate}}
\newcommand{\een}{\end{enumerate}}

\theoremstyle{definition}
\newtheorem{theorem}{Theorem}
\theoremstyle{definition}
\newtheorem{corollary}{Corollary}
\theoremstyle{definition}
\newtheorem{lemma}{Lemma}

\theoremstyle{definition}
\newtheorem{definition}{Definition}

\usepackage{array}
\usepackage{amsmath}

\newcommand{\bea}{\begin{eqnarray}}
\newcommand{\eea}{\end{eqnarray}}
\newcommand{\be}{\begin{equation}}
\newcommand{\ee}{\end{equation}}
\newcommand{\ba}{\begin{equation}\begin{aligned}}
\newcommand{\ea}{\end{aligned}\end{equation}}

\newcommand{\bee}{\begin{enumerate}}
\newcommand{\eee}{\end{enumerate}}

\def\be{\begin{equation}}
\def\ee{\end{equation}}

\newcommand{\herm}{{\rm Herm}}
\newcommand{\uD}{\underline{\mathbf{D}}}

\newcommand{\mb}{\mathfrak{B}}
\newcommand{\md}{\mathfrak{D}}
\newcommand{\mf}{\mathfrak{F}}

\newcommand{\mr}{\mathring}

\newcommand{\mM}{\mathcal{M}}

\newcommand{\D}{\mathbf{D}}

\newcommand{\lr}{\rangle\langle}
\newcommand{\la}{\langle}
\newcommand{\ra}{\rangle}
\newcommand{\tr}{{\rm Tr}}

\newcommand{\eps}{\varepsilon}

\newcommand{\mbb}[1]{\mathbb{#1}}

\newcommand{\bra}[1]{\langle #1|}
\newcommand{\ket}[1]{|#1\rangle}

\newcommand{\eqdef}{\coloneqq}

\def\p{\mathbf{p}}

\def\e{\mathbf{e}}

\def\u{\mathbf{u}}

\def\0{\mathbf{0}}

\def\tA{\tilde{A}}

\def\tD{\tilde{D}}

\def\trho{\tilde{\rho}}

\def\ttau{\tilde{\tau}}

\def\eff{{\rm eff}}

\newcommand{\GG}[1]{\rm \textcolor{red}{ #1}{\color{red} \to}}

\newcommand{\Gg}[1]{
  \text{\fcolorbox{black}{red!20}{\textcolor{black}{\scriptsize$#1$}}}{\color{red} \xrightarrow{\hspace{0.1cm}\;\hspace{0.1cm}}}
}

\begin{document}

\title{Induced Quantum Divergence: A New Lens on Communication and Source Coding}
\author{Gilad Gour}
\affiliation{Technion - Israel Institute of Technology, Faculty of Mathematics, Haifa 3200003, Israel}
	
	\date{\today}

\begin{abstract}	
	This paper introduces the induced divergence, a new quantum divergence measure that replaces the hypothesis testing divergence in position-based decoding, simplifying the analysis of quantum communication and state redistribution while yielding tighter achievability bounds. Derived from a parent quantum relative entropy, it retains key properties such as data processing inequality and L\"owner monotonicity. Like the hypothesis testing divergence, it depends on a smoothing parameter and interpolates between the parent relative entropy (as the smoothing parameter approaches one) and the min-relative entropy (as it approaches zero), the latter holding when applied to the sandwiched R\'enyi relative entropy of order $\alpha\in[0,2]$. This framework refines the position-based decoding lemma, extending its applicability to a broader class of states and improving decoding success probabilities. Two key applications are considered: classical communication over quantum channels, where the induced divergence improves lower bounds on the distillable communication rate, and quantum state redistribution, where it leads to sharper bounds on communication costs. These results provide new insights into fundamental single-shot quantum information protocols and enhance existing analytical techniques.

	\end{abstract}
	
	\maketitle


\section{Introduction}

Understanding the fundamental limits of quantum communication and information processing is a central objective in quantum information science~\cite{Hayashi2006,Wilde2013,Tomamichel2015,Watrous2018,KW2024,Gour2025}. Many quantum protocols, whether in communication or source coding, require analyzing how accurately a quantum state can be transferred or reconstructed under constraints on operations and available resources. A key challenge in these tasks is identifying divergence measures that provide both operational significance and mathematically tractable bounds on achievable rates, such as distillable rates or resource costs of quantum information processing (QIP) tasks.

A widely used divergence in single-shot quantum information theory is the hypothesis testing divergence, which naturally arises in both classical and quantum hypothesis testing. It is computationally efficient, admits an operational interpretation, and plays a central role in achievability results for quantum protocols, including its key role in the recent proof of the generalized quantum Stein's lemma~\cite{HY2024}. Moreover, it has numerous applications in quantum Shannon theory and quantum resource theories (see for example the books~\cite{Hayashi2006,Wilde2013,Tomamichel2015,Watrous2018,KW2024,Gour2025} and references therein). However, this paper demonstrates that in certain scenarios, even tighter bounds can be achieved by replacing the hypothesis testing divergence with a new measure: the induced divergence.

The induced divergence is derived from general quantum relative entropies and satisfies fundamental properties, including the data processing inequality. It is a smoothed divergence, defined with respect to a smoothing parameter $\eps\in[0,1]$, and smoothly interpolates between two key divergences. In the limit $\eps\to 1^-$, it converges to its parent relative entropy, while for small $\eps$, it closely approximates the hypothesis testing divergence -- particularly when based on the sandwiched R\'enyi relative entropy for $\alpha \in [0,2]$. This makes it especially effective in single-shot domain. In particular, we will show that the induced divergence of the collision relative entropy (i.e. the sandwiched R\'enyi relative entropy of order $\alpha=2$) has significant applications in both communication and source coding.

A major application of this framework is an enhanced position-based decoding lemma, which extends its applicability to a broader class of quantum states and improves the probability of successful decoding. The position-based decoding lemma~\cite{AJW2018,AJW2019} has emerged as a versatile tool for analyzing quantum communication and state redistribution protocols~\cite{AJ2022,ABS+2023}. Our refinements strengthen achievability results in these areas, leading to improved bounds for classical communication over quantum channels, where the induced divergence provides tighter lower bounds on the distillable communication rate. Similarly, in quantum state redistribution, we obtain a sharper characterization of the required quantum communication cost, surpassing previous results based on the hypothesis testing divergence.

Beyond these specific applications, the induced divergence serves as a versatile tool with the potential to refine a wide range of quantum information protocols. It is also an intriguing mathematical construct, worthy of study in its own right. While the hypothesis testing divergence is both computationally efficient and operationally meaningful, the induced divergence provides an alternative that yields even tighter bounds in certain key quantum information tasks. This work, therefore, advances our understanding of quantum communication limits.

The remainder of the paper is organized as follows. Section II reviews relevant quantum divergences, including the R\'enyi and hypothesis testing divergences. Section III introduces the induced divergence, establishing its definition and key properties. In Section IV, we present the enhanced position-based decoding lemma and its implications. Section V applies these results to classical communication over quantum channels, while Section VI analyzes quantum state redistribution. Finally, Section VII concludes with open questions and future research directions.

\section{Preliminaries}

\subsection{Notations}

The letters $A$, $B$, and $R$ will be used to denote both quantum systems (or registers) and their corresponding Hilbert spaces, with $|A|$ representing the dimension of the Hilbert space associated with system $A$. Throughout, we consider only finite-dimensional Hilbert spaces. The letters $X$, $Y$, and $Z$ will be used to describe classical systems. To indicate a replica of a system, we use a tilde symbol above its label; for instance, $\tA$ denotes a replica of $A$, implying $|\tA| = |A|$. 

The set of all positive operators acting on system $A$ is denoted by $\pos(A)$, and the set of all density operators in $\pos(A)$ is denoted by $\md(A)$, with its subset of pure states represented by $\pure(A)$. We will also denote by $\prob(m)$ the set of all probability vectors in $\mbb{R}^m$, and by $\eff(A)$ the set of all effects in $\pos(A)$ (i.e., all operators $\Lambda\in\pos(A)$ with $\Lambda\leq I^A$). 

The set of all quantum channels, i.e., completely positive and trace-preserving (CPTP) maps from system $A$ to system $B$, is denoted by $\cptp(A \to B)$. Elements of $\cptp(A \to B)$ are represented by calligraphic letters such as $\mE$, $\mF$, and $\mN$. We often interchange $\mN^{A\to B}$ and $\mN^{\tA\to B}$, as when $\mN$ acts on a state $\rho\in\md(A\tA)$, we write the resulting state as $\omega^{AB}\eqdef\mN^{\tA\to B}(\rho^{A\tA})$ to avoid a tilde on $\omega^{AB}$. Here, $\tA$ simply serves as a replica of $A$.

We use in this paper the trace distance as the primary metric. Hence, we define a ball of radius $\eps>0$ around a state $\rho\in\md(A)$ as:
\be
\mb^\eps(\rho)\eqdef\left\{\sigma\in\md(A)\;:\;\frac12\|\rho-\sigma\|_1\leq\eps\right\}\;.
\ee
We say that sigma is $\eps$-close to $\rho$ and write $\sigma\approx_\eps\rho$ if $\sigma\in\mb^\eps(\rho)$.
The purified distance between two density matrices $\rho,\sigma\in\md(A)$ is defined as
\be
P(\rho,\sigma)\eqdef\sqrt{1-F^2(\rho,\sigma)}
\ee
where $F(\rho,\sigma)\eqdef\|\sqrt{\rho}\sqrt{\sigma}\|_1$ is the fidelity.

Moreover, for every $\Lambda\in\pos(A)$ we use the notation $\spec(\Lambda)$ to denote the spectrum of $\Lambda$ (that is, the set of distinct eigenvalues of $\Lambda$).

\subsection{The Sandwiched R\'enyi Relative Entropy}

In this paper we primarily work with the \emph{sandwiched R\'enyi relative entropy}.  The sandwiched R\'enyi relative entropy of order $\alpha \in [0,\infty]$ is defined for any quantum system $A$ and for $\rho, \sigma \in \pos(A)$ as~\cite{MDS+2013,WWY2014,Matsumoto2018b,GT2020}
\be\nonumber
D_{\alpha}(\rho\|\sigma)=
	\begin{cases}
	\substack{\frac1{\alpha-1}\log Q_\alpha(\rho\|\sigma)\\}&\substack{\text{if }\frac12\leq\alpha<1\text{ and }\rho\not\perp\sigma, \\\text{ or }\alpha>1 \text{ and } \rho\ll\sigma\\}\\
	\substack{\frac1{\alpha-1}\log Q_{1-\alpha}(\sigma\|\rho)\\}&\substack{\text{if }0\leq \alpha<\frac12\text{ and }\rho\not\perp\sigma\\}\\
	\substack{\infty\\}&\substack{\text{otherwise}\\}
	\end{cases}
\ee
Here, $\rho \ll \sigma$ means that the support of $\rho$ is a subspace of the support of $\sigma$, and $\rho \not\perp \sigma$ means $\tr[\rho \sigma] \neq 0$. The quantity $Q_\alpha(\rho\|\sigma)$ is defined as
\be
Q_\alpha(\rho\|\sigma)\eqdef\tr\left(\sigma^{\frac{1-\alpha}{2\alpha}}\rho\sigma^{\frac{1-\alpha}{2\alpha}}\right)^\alpha\;.
\ee

The sandwiched Rényi relative entropy is the smallest quantum divergence that reduces to the classical R\'enyi relative entropy in the commutative case.
It admits special cases for $\alpha=0$, $\alpha=1$, and $\alpha=\infty$, which are understood in terms of limits:
\ben
\item For $\alpha=0$, it corresponds to the min relative entropy $D_{\min}$, defined as:
\be 
D_{\min}(\rho\|\sigma)\eqdef-\log\tr\left[\sigma\Pi_\rho\right]\;, 
\ee
where $\Pi_\rho$ is the projection onto the support of $\rho$.
\item For $\alpha=1$, it reduces to the Umegaki relative entropy:
\be 
D(\rho\|\sigma)\eqdef\tr[\rho\log\rho]-\tr[\rho\log\sigma]\;.
\ee
\item 
For $\alpha=\infty$, it gives the max relative entropy:
\be
D_{\max}(\rho\|\sigma)\eqdef\inf_{t\in\mbb{R}_+}\big\{\log t\;:\;t\sigma\geq\rho\big\}\;.
\ee
\een

A useful property of $Q_\alpha$ (for all $\alpha\in[0,\infty]$) that we will employ in this paper is its behavior under direct sums. Specifically, consider two cq-states $\rho, \sigma \in \md(XA)$ of the form:
\be\label{r}
\rho^{XA}\eqdef\sum_{x\in[k]}p_x|x\lr x|^X\otimes\rho_x^A
\ee
and
\be\label{s}
\sigma^{XA}\eqdef\sum_{x\in[k]}p_x|x\lr x|^X\otimes\sigma_x^A\;.
\ee
where $\rho^X = \sigma^X$. For such states, $Q_\alpha$ satisfies the following \emph{direct sum property}:
\be\label{dsp}
Q_\alpha\left(\rho^{XA}\big\|\sigma^{XA}\right)=\sum_{x\in[k]}p_xQ_\alpha\left(\rho^A_x\big\|\sigma_x^A\right)\;.
\ee
This property will be useful to some of the results discussed in this paper.

\subsection{The Collision Relative Entropy}

The sandwiched R\'enyi relative entropy of order $\alpha = 2$, often referred to as the \emph{collision relative entropy}~\cite{BG2014,Gour2025b}, will also play a central role in several applications discussed in this paper. Its name is derived from the concept of collision entropy, which is closely tied to the \emph{collision probability} in probability theory. For all $\rho, \sigma \in \md(A)$, it is defined as
\be
D_2(\rho\|\sigma)\eqdef\log Q_2(\rho\|\sigma)
\ee
where
\ba
Q_2(\rho\|\sigma)&\eqdef\tr\left[\left(\sigma^{-\frac14}\rho\sigma^{-\frac14}\right)^2\right]\\
&=\tr\left[\rho\sigma^{-\frac12}\rho\sigma^{-\frac12}\right]\;.
\ea
This quantity has a particularly simple form due to the quadratic dependence of $Q_2$ on $\rho$. Recently, this property was harnessed to obtain an equality-based version of the convex split lemma~\cite{Gour2025b} and consequently shown to play a key role in certain applications that make use of the convex split lemma~\cite{ADJ2017}, such as quantum state merging~\cite{HOW2005,Berta2009} and state splitting~\cite{ADHW2009,BCR2011}.

The collision relative entropy can be connected to the purified distance. Specifically, since $D_{1/2}$ is bounded above by $D_2$, it follows that for every $\rho, \sigma \in \md(A)$
\be
D_{2}(\rho\|\sigma)\geq-\log\left(1-P^2(\rho,\sigma)\right)\;.
\ee
This relation can be expressed as
\be
P^2(\rho,\sigma)\leq 1-\frac{1}{Q_2(\rho\|\sigma)}\;.
\ee

\subsection{The Hypothesis Testing and Information Spectrum Divergence}

For any $\eps\in[0,1)$, $\rho\in\md(A)$, and $\sigma\in\pos(A)$, the hypothesis testing divergence is defined as:
\be\label{htd}
D_{\min}^\eps(\rho\|\sigma)\eqdef-\log\min_{\Lambda\in\eff(A)}\Big\{\tr[\Lambda\sigma]\;:\;\tr[\rho\Lambda]\geq1-\eps\Big\}\;.
\ee
This divergence plays a central role in many QIP tasks. For example, the quantum Stein's lemma can be expressed in terms of the hypothesis testing divergence as follows: for all $\rho,\sigma\in\md(A)$ and $\eps\in(0,1)$,
\be\label{qsl}
\lim_{n\to\infty}\frac1n{D}^\eps_H\left(\rho^{\otimes n}\big\|\sigma^{\otimes n}\right)=D(\rho\|\sigma)\;,
\ee
where $D(\rho\|\sigma)\eqdef\tr[\rho\log\rho]-\tr[\rho\log\sigma]$ is the Umegaki relative entropy.

In this paper, we also consider two variants of the information spectrum divergence, introduced in~\cite{DL2015}. For any $\rho,\sigma\in\md(A)$ and $\eps\in(0,1)$, they are defined as:
\ba\label{iss}
&\underline{D}_s^\eps(\rho\|\sigma)\eqdef\sup_{\lambda\in\mbb{R}}\{\lambda\;:\;\tr(\rho-2^\lambda\sigma)_+\geq1-\eps\}\\
&\overline{D}_s^\eps(\rho\|\sigma)\eqdef\inf_{\lambda\in\mbb{R}}\{\lambda\;:\;\tr(\rho-2^\lambda\sigma)_+\leq\eps\}
\ea
These divergences are variants of the definition originally introduced in~\cite{TH2013}.

Since the optimization conditions for $\underline{D}_s^\eps(\rho\|\sigma)$ and $\overline{D}_s^\eps(\rho\|\sigma)$ are attained when $\tr(\rho-2^\lambda\sigma)_+=1-\eps$ and $\tr(\rho-2^\lambda\sigma)_+=\eps$, respectively, it follows that 
\be
\underline{D}_s^\eps(\rho\|\sigma)=\overline{D}_s^{1-\eps}(\rho\|\sigma)
\ee 
(see~\cite{DL2015}). Consequently, it is sufficient to consider only one of them. Here, we adopt the notation introduced in~\cite{RLD2025}:
\be
\tilde{D}_{\max}^\eps(\rho\|\sigma)\eqdef\overline{D}_s^\eps(\rho\|\sigma)\;.
\ee
This notation reflects the fact that $\overline{D}_s^\eps$ is a smoothed variant of $D_{\max}$; specifically, for $\eps=0$, $\overline{D}_s^0=D_{\max}$ (see~\cite{RLD2025} for further details). Moreover, it was shown in~\cite{RLD2025} that $\tilde{D}_{\max}^\eps$ is related to the hypothesis testing divergence through the following relations:
\ba\label{is}
\tilde{D}_{\max}^\eps(\rho\|\sigma)=\sup_{\delta\in(\eps,1]}\left\{D_{H}^{1-\delta}(\rho\|\sigma)+\log(\delta-\eps)\right\}\\
D_{H}^{1-\eps}(\rho\|\sigma)=\inf_{\delta\in[0,\eps)}\left\{\tilde{D}_{\max}^{\delta}(\rho\|\sigma)-\log(\eps-\delta)\right\}
\ea

\subsection{R\'enyi Mutual Information}

Every relative entropy (i.e., additive quantum divergence) can be employed to define other entropic functions. In the context of the applications considered in this paper, we frequently encounter the mutual information. For every $\alpha\in[0,\infty]$, we define the $\alpha$-mutual information of a bipartite state as follows:
\be\label{alpha} 
I_\alpha(A:B)_\rho\eqdef\min_{\sigma\in\md(B)}D_\alpha\left(\rho^{AB}\big\|\rho^A\otimes\sigma^B\right)\;. 
\ee 
While alternative definitions of mutual information have been proposed in the literature (see, for example,~\cite{CBR2014} and references therein), often with various operational interpretations, we adopt the above definition as it is the most relevant to the applications considered in this paper.

In particular, we focus on three notable special cases of the mutual information:
\ben \item \textbf{The case $\alpha=1$:} Here, $I_1(A:B)_\rho$, denoted simply as $I(A:B)_\rho$, is expressed as
\be 
I(A:B)_\rho=D\left(\rho^{AB}\big\|\rho^A\otimes\rho^B\right)\;, 
\ee 
(in this case the minimization over $\sigma\in\md(B)$ is achieved for $\sigma^B=\rho^B$).
\item \textbf{The case $\alpha=2$:} This case is given by
\be\label{i2}
I_2(A:B)_\rho\eqdef\min_{\sigma\in\md(B)}\log Q_2\left(\rho^{AB}\big\|\rho^A\otimes\sigma^B\right)\;. 
\ee 
\item \textbf{The case $\alpha=\infty$:} This is expressed in terms of the max-relative entropy as
\be 
I_{\max}(A:B)_\rho\eqdef\min_{\sigma\in\md(B)}D_{\max}\left(\rho^{AB}\big\|\rho^A\otimes\sigma^B\right)\;. 
\ee 
\een
We also consider the smoothed version of the R\'enyi mutual information, defined for all $\eps\in(0,1)$ and $\rho\in\md(AB)$ as:
\be\label{srmi}
I_{\alpha}^\eps(A:B)_\rho\eqdef\min_{\rho'\in\mb^\eps(\rho)}I_{\alpha}(A:B)_{\rho'}\;. 
\ee

\subsection{Measured R\'enyi Relative Entropy} 

The measured R\'enyi relative entropy of order $\alpha\in[0,\infty]$ is defined for every $\rho\in\md(A)$ and all $\sigma\in\pos(A)$ as
\be\label{measured}
D^{\mbb{M}}_\alpha(\rho\|\sigma)\eqdef\sup_{X,\mE\in\cptp(A\to X)}D_\alpha\left(\mE(\rho)\big\|\mE(\sigma)\right)
\ee
where the supremum is over all finite dimensional classical systems $X$ and all quantum channels in $\cptp(A\to X)$. Due to the DPI we have for all $\rho\in\md(A)$ and all $\sigma\in\pos(A)$
\be
D_\alpha(\rho\|\sigma)\geq D^{\mbb{M}}_\alpha(\rho\|\sigma)\;.
\ee
Moreover, in~\cite{BFT2017} it was shown that equality holds iff $[\rho,\sigma]=0$. 

For $\alpha\in[0,2]$ the sandwiched relative entropy satisfies~\cite{Tomamichel2015}:
\be
D_\alpha\left(\mP_\sigma(\rho)\big\|\sigma\right)\geq D_\alpha(\rho\|\sigma)+\log|\spec(\sigma)|\;,
\ee
where $\mP_\sigma$ is the pinching channel. We therefore also have
\be\label{measured2}
D_\alpha^{\mbb{M}}(\rho\|\sigma)\geq D_\alpha(\rho\|\sigma)+\log|\spec(\sigma)|\;.
\ee
Additional tight relations between various smoothed relative entropies can be found in~\cite{RLD2025}.

\subsection{Extensions to the Channel Domain}

A superchannel is a completely positive linear map that transforms quantum channels into quantum channels. It has been shown~\cite{CDP2008,Gour2019} that any superchannel $\Theta$ mapping elements of $\cptp(A\to B)$ to elements of $\cptp(A'\to B')$ can be realized via a pre-processing map $\mE\in\cptp(A'\to RA)$ and a post-processing map $\mF\in\cptp(RB\to B')$, such that for every $\mN\in\cptp(A\to B)$,
\be
\Theta[\mN]=\mF^{RB\to B'}\circ\mN^{A\to B}\circ\mE^{A'\to RA}\;.
\ee
Using this property, we define a channel divergence as a map $\D$ that assigns a real value to pairs of quantum channels in $\cptp(A\to B)$ and satisfies the generalized data-processing inequality (DPI): for every superchannel $\Theta$ and channels $\mN,\mM\in\cptp(A\to B)$,
\be
\D\left(\Theta[\mN]\big\|\Theta[\mM]\right)\leq\D(\mN\|\mM)\;.
\ee
We consider only channel divergences that are normalized, meaning that $\D(\mN\|\mN) = 0$.

Channel divergences naturally extend state divergences, since when $|A|=1$, the channels $\mN$ and $\mM$ reduce to quantum states, and $\Theta$ corresponds to a quantum channel. There are several standard methods for extending quantum divergences to channels (see, e.g.,~\cite{Gour2021} and references therein), but here we focus on one particular extension. Given a quantum divergence $\D$ defined on quantum states, we extend it to channels in $\cptp(A\to B)$ (for any finite-dimensional systems $A$ and $B$) as:\footnote{For simplicity, we use the same notation $\D$ for both quantum and channel divergences.}
\be\label{cd}
\D(\mN\|\mM)\eqdef\sup_{\rho\in\md(RA)}\D\left(\mN^{A\to B}\left(\rho^{RA}\right)\big\|\mM^{A\to B}\left(\rho^{RA}\right)\right)
\ee
where the supremum is taken over all density matrices in $\md(RA)$ and all systems $R$. However, it can be shown that it suffices to restrict to $|R|=|A|$ and take $\rho^{RA}$ to be pure. 

This approach can also be used to extend other entropic functions to the channel domain. More generally, a framework for extending resource monotones from one domain (such as quantum states) to another domain (such as quantum channels) has been developed in~\cite{GT2020}. As an example, the $\alpha$-mutual information, as defined in~\eqref{alpha}, can be extended to a channel mutual information as follows. Let $\mN \in \cptp(A\to B)$ and $\alpha\in[0,\infty]$. We define the $\alpha$-mutual information of the channel $\mN$ as:
\be
I_\alpha(A:B)_\mN\eqdef\sup_{\psi\in\pure(A\tA)}I_\alpha(A:B)_{\omega_\psi}\;,
\ee
where $\omega^{AB}_\psi\eqdef\mN^{\tA\to B}\big(\psi^{A\tA}\big)$.

In our application to classical communication over a quantum channel, we will encounter both channel divergences and mutual information, but with $A=X$ as a classical system. In this case, every $\psi\in\pure(A\tA)$ can be replaced by its Schmidt probability vector $\p\in\prob(k)$, leading to
\be\label{sp}
\omega^{XB}_\psi=\sigma^{XB}_\p\eqdef\sum_{x\in[k]}p_x|x\lr x|^X\otimes\sigma_x^B\;,
\ee
where $\sigma_x^B\eqdef\mN(|x\lr x|)$. 

If the state divergence $\D$ satisfies the direct sum property~\eqref{dsp}, then its channel extension in~\eqref{cd} simplifies to
\be
\D(\mN \| \mM)=\sup_{x\in[k]} \D\left(\mN(|x\lr x|) \big\| \mM\big(|x\lr x|)\right)\;.
\ee
Similarly, the $\alpha$-mutual information for such a channel is given by 
\be
I_\alpha(X:B)_\mN=\sup_{\p\in\prob(m)}I_\alpha(X:B)_{\sigma_\p}
\ee
where $\sigma_\p^{XB}$ is defined in~\eqref{sp}.

\section{The Induced Divergence}

In this section, we introduce a novel divergence, referred to as the ``induced divergence", which can be constructed from any quantum relative entropy and plays a central role in this paper. For certain sandwiched Rényi relative entropies, the induced divergence bears similarities to the hypothesis testing divergence, as it can be interpreted as a smoothed variant of the min-relative entropy. However, it also exhibits notable differences, establishing itself as a unique and versatile tool. This newly defined divergence is particularly powerful for analyzing single-shot QIP tasks, leading to significant improvements in several established results. Its unique properties and operational relevance will be highlighted in the following sections as we delve into its diverse applications.

Before presenting the definition of the induced divergence, we first discuss the extension of quantum relative entropies to larger domains, an essential step for defining the induced divergence.

\subsection{Relative Entropies with Extended Domain}

A quantum divergence $\D$ is a function that maps pairs of quantum states in $\md(A)$ to the real line, satisfying the data processing inequality (DPI). Additionally, a quantum divergence is called a relative entropy if it meets two further conditions: additivity under tensor products and the following normalization condition~\cite{GT2020,GT2021}:
\be\label{1}
\D\left(|0\lr 0|\Big\|\frac12 I_2\right)=1\;,
\ee
where it is important to note that this condition is not symmetric.

In~\cite{GT2020,GT2021}, it was shown that these conditions imply the following: For every system $A$ with dimension $m \eqdef |A|$, we have
\be\label{m2}
\D\left(|0\lr 0|^A\big\|\u^A\right)=\log (m)\;,
\ee
where $\u^A$ is the maximally mixed state in $\md(A)$.
More generally, it was shown that for any diagonal density matrix $\rho = \sum_{x \in [m]} p_x |x\lr x|$, we have
\be\label{key}
\D\left(|x\lr x|\big\|\rho\right)=-\log (p_x)\;.
\ee

We extend the definition from~\cite{GT2020,GT2021} to allow the second argument of $\D$ to include not only density matrices but also positive semi-definite operators. Explicitly, we consider a function
\be\label{0}
\D:\bigcup_{A}\md(A)\times\pos(A)\to\mbb{R}
\ee
where the union is taken over all finite-dimensional Hilbert spaces. That is, $\D$ is defined on pairs $(\rho, \sigma)$, where $\rho$ is a density matrix in $\md(A)$, and $\sigma$ is a positive semi-definite operator in $\pos(A)$.

We say that $\D$ is a quantum divergence if it satisfies the extended DPI: for all $\rho \in \md(A)$, $\sigma \in \pos(A)$, and $\mE \in \cptp(A \to B)$,
\be\label{dpi}
\D\left(\mE(\rho)\big\|\mE(\sigma)\right)\leq\D(\rho\|\sigma)\;.
\ee
Furthermore, $\D$ is said to be additive if, for all $\rho \in \md(A)$, $\sigma \in \pos(A)$, $\rho' \in \md(A')$, and $\sigma' \in \pos(A')$,
\be\label{additivity}
\D(\rho\otimes\rho'\|\sigma\otimes\sigma')=\D(\rho\|\sigma)+\D(\rho'\|\sigma')\;.
\ee

Since we are extending the action of $\D$ beyond density matrices in the second argument, it will  convenient to introduce another normalization condition. To see why, let $\D$, as given in~\eqref{0}, be an additive quantum divergence satisfying the DPI in the form~\eqref{dpi},
the additivity in the form~\eqref{additivity}, and the
normalization condition~\eqref{1}. Now, fix a non-zero $r\in\mbb{R}$, and observe that the function
\be
\D'(\rho\|\sigma)\eqdef\D(\rho\|\sigma)+r\log\tr[\sigma]
\ee
is also an additive quantum divergence satisfying the DPI in the form~\eqref{dpi},
the additivity in the form~\eqref{additivity}, and the
normalization condition~\eqref{1}. This additional freedom emerges since we allow the second argument, $\sigma$, to have arbitrary positive trace. To eliminate this freedom, it is sufficient to consider the function 
\be
t\mapsto\D(1\|t)\quad\quad\forall\;t\in\mbb{R}_+\;,
\ee
where the number $1$ inside $\D(1\|t)$ is viewed as a (trivial) density matrix in $\md(\mbb{C})$, and the number $t\geq 0$ as a positive semi-definite operator in $\pos(\mbb{C})=\mbb{R}_+$.
Due to the additivity of $\D$ we have $\D(1\|1)=0$~\cite{GT2021}, and we choose the normalization 
\be\label{2}
\D(1\|2)=-1\;.
\ee
As we will see shortly,
this normalization condition implies that:
\be
\D\left(|0\lr 0|^A\big\|I^A\right)=0\;,
\ee
for every finite-dimensional system $A$.
This condition is satisfied by all relative entropies studied in the literature. Importantly, it ensures that the map $\rho \mapsto -\D(\rho \| I)$ can be interpreted as an entropy function.
To summarize:

\bmyd\label{def1}
A function $\D$, as defined in~\eqref{0}, is a quantum relative entropy if it satisfies the DPI~\eqref{dpi}, additivity~\eqref{additivity}, and the normalization conditions~\eqref{1} and~\eqref{2}.
\emyd

Classically, any relative entropy that is continuous in its second argument can be expressed as a convex combination of the R\'enyi relative entropies~\cite{MPS+2021}. Consequently, we will restrict our attention here to quantum relative entropies that are continuous in their second argument.

The extension of relative entropies to the quantum domain is not unique, leading to a richer landscape of quantum relative entropies. Nevertheless, as demonstrated in the following lemma, some classical properties carry over to the quantum setting.

Let $\D$ be a quantum relative entropy, $t\in\mbb{R}_+$, $\rho\in\md(A)$, and $\sigma, \sigma' \in \pos(A)$. Then:
\bmyl\label{lowner}$\;$
\ben
\item \textbf{Scaling property:}
\be\label{scale}
\D(\rho\|t\sigma) = \D(\rho\|\sigma) - \log t.
\ee
\item \textbf{L\"owner monotonicity:} If $\sigma' \geq \sigma$, then
\be\label{lm}
\D(\rho\|\sigma) \geq \D(\rho\|\sigma').
\ee
\een
\emyl

\begin{proof}
For the scaling property, note first that since $\rho=\rho\otimes 1$ and $t\sigma = \sigma \otimes t$, we get that
\ba
\D(\rho\|t\sigma)&=\D\left(\rho\otimes 1\big\|\sigma\otimes t\right)\\
\GG{Additivity}&=\D(\rho\|\sigma)+f(t)
\ea
where $f(t) \eqdef \D(1\|t)$. Observe that for any $s, t \in \mbb{R}_+$,
\ba
f(ts)=\D(1\|ts)&=\D(1\otimes 1\|t\otimes s)\\
\GG{Additivity}&=f(t)+f(s)\;.
\ea
Moreover, by assumption, $f(2)=-1$ and $f$ is continuous (since we assume that $\D$ is continuous in its second argument). The only additive function with these properties is $f(t)=-\log(t)$ (here $\log=\log_2$). 
This complete the proof of~\eqref{scale}.

For L\"owner monotonicity, let $\omega \eqdef \sigma' - \sigma$. By assumption, $\omega \geq 0$. The case $\omega = 0$ is trivial, so we assume $\tr[\omega] > 0$ and define
\be
t\eqdef\frac{\tr[\omega]}{\tr[\omega]+\tr[\sigma]}\in(0,1)\;.
\ee
Then, from~\eqref{key} we get that
\ba
&{\D}(\rho\|\sigma)-{\log} (t)\\
&=\D\left(\rho\|\sigma\right)+\D\left(|0\lr 0|\Big\|t|0\lr 0|+(1-t)|1\lr 1|\right)\\
&=\D\left(\rho\otimes |0\lr 0|\Big\|t\sigma\otimes|0\lr 0|+(1-t)\sigma\otimes|1\lr 1|\right)\;.
\ea
Let $\mE \in \cptp(AX \to A)$ be a quantum channel that models the process of measuring $X$. If the outcome is $\ket{0}\bra{0}$, the channel leaves system $A$ intact; otherwise, it replaces the state of system $A$ with the normalized state $\omega /\tr[\omega]$. Then, by the DPI,
\ba
&{\D}(\rho\|\sigma)-{\log} t\\
&\geq\D\left(\mE(\rho\otimes |0\lr 0|)\Big\|t\mE(\sigma\otimes|0\lr 0|)+(1-t)\mE(\sigma\otimes|1\lr 1|)\right)\\
&=\D\left(\rho\Big\|t\sigma+(1-t)\tr[\sigma]\frac{\omega}{\tr[\omega]}\right)\\
&={\D(\rho\|\sigma+\omega)-\log t}\;,
\ea
where in the last line we used the scaling property~\eqref{scale} after observing that from its definition $t=(1-t)\frac{\tr[\sigma]}{\tr[\omega]}$. Thus,
\ba
\D(\rho\|\sigma)&\geq \D(\rho\|\sigma+\omega)\\
&=\D(\rho\|\sigma')\;.
\ea
This completes the proof.
\end{proof}

Beyond these foundational properties, the sandwiched R\'enyi relative entropy satisfies the following inequality.

Let $\rho,\sigma\in\pos(A)$, $(\rho-\sigma)_+$ denotes the positive part of $\rho-\sigma$, and $\alpha\in[0,2]$. Then:
\bmyl\label{cool0}
\be\label{141}
Q_\alpha\left(\rho\big\|\rho+\sigma\right)\geq\tr(\rho-\sigma)_+\;.
\ee 
\emyl

\begin{proof}
The commutative case, where $[\rho,\sigma]=0$, follows directly from the inequality $p^\alpha(p+q)^{1-\alpha}\geq (p-q)_+$, applied to each eigenvalue of $\rho^\alpha(\rho+\sigma)^{1-\alpha}$. Thus, it remains to prove the inequality for the non-commutative case.
Since $D_\alpha$ satisfies the DPI, for any classical system $X$ and any POVM channel $\mM\in\cptp(A\to X)$, we have
\ba
Q_\alpha\left(\rho\|\rho+\sigma\right)&\geq Q_\alpha\left(\mM(\rho)\big\|\mM(\rho+\sigma)\right)\\
\GG{\small{\text{Commutative Case}}}&\geq\tr\big(\mM(\rho-\sigma)\big)_+\;.
\ea
Taking $|X|=2$ and $\mM(\omega)=\tr[\Pi_+\omega]|0\lr 0|+\tr[\Pi_-\omega]|1\lr 1|$ with $\Pi_{\pm}\in\pos(A)$ being the projections to the positive and negative eigenspaces of $\rho-\sigma$ we get that 
$\tr\big(\mM(\rho-\sigma)\big)_+=\tr(\rho-\sigma)_+$. This completes the proof.
\end{proof}

The inequality in~\eqref{141} can be applied to provide a straightforward proof of a related inequality established in~\cite{Cheng2023}. Specifically, recall that the minimum of two real numbers $p, q \in \mbb{R}$ can be expressed as
\be
\min\{p,q\}=\frac12(p+q-|p-q|)\;.
\ee
While, in general, the minimum between two Hermitian operators does not exist, we can use this formula to define an extension for all $\rho, \sigma \in \herm(A)$:
\ba
\rho\land \sigma\eqdef\frac12(\rho+\sigma-|\rho-\sigma|)\;.
\ea
In~\cite{ACM+2007} it was shown that for $\rho, \sigma \geq 0$, the inequality $\tr[\rho \land \sigma] \leq \tr[\rho^s \sigma^{1-s}]$ holds for all $s \in [0, 1]$. Here, we show that the inequality in~\eqref{141} can be used to provide a simple proof for the lower bound found in~\cite{Cheng2023} for $\tr[\rho \land \sigma]$. This lower bound generalizes the classical inequality $\min\{p, q\} \geq \frac{pq}{p + q}$.

Let $\rho,\sigma\in\pos(A)$ and set $\Lambda\eqdef \rho+\sigma$. Then:
\bmyc\cite{Cheng2023}
\be\label{mainineq}
\tr[\rho\land\sigma]\geq\tr\left[\rho\Lambda^{-1/2}\sigma\Lambda^{-1/2}\right] \;.
\ee
\emyc

\begin{proof}
The proof follows from the simple observation that
\ba
\tr\left[\rho\Lambda^{-1/2}\sigma\Lambda^{-1/2}\right]&=\tr[\rho]-Q_2(\rho\|\Lambda)\\
\GG{\eqref{141}}&\leq \tr[\rho]-\tr(\rho-\sigma)_+\\
&=\tr[\rho\land\sigma]\;.
\ea
 This completes the proof.
\end{proof}

\subsection{The Induced Divergence: Definition and Basic Properties}

We now introduce a novel divergence, called the \emph{induced divergence}, which is a type of smoothed divergence and derive some of its properties. Similar to $D_H^\eps$ and $\tD_{\max}^\eps$ this divergence is not normalized, so we will also define its normalized version. 

Let $\eps \in (0,1)$ and let $\D$ be a quantum relative entropy as defined in Definition~\ref{def1}, which is continuous in its second argument. The induced divergence $\mr{\D}^\eps$ and its normalized version $\mr{\uD}^\eps$ are defined for all $\rho\in\md(A)$ and $\sigma \in \pos(A)$ as follows:
\bmyd
\ba\label{idn}
&\mr{\D}^\eps(\rho\|\sigma)\eqdef\sup_{\lambda\in\mbb{R}}\big\{\lambda\;:\;\D\left(\rho\|\rho+2^\lambda\sigma\right)\geq \log(1-\eps)\big\}\\
&\mr{\uD}^\eps(\rho\|\sigma)\eqdef\mr{\D}^\eps(\rho\|\sigma)+\log\left(\frac{1-\eps}\eps\right)\;.
\ea
\emyd

The term $\log\left(\frac{1-\eps}{\eps}\right)$ in the definition of the normalized induced divergence ensures that
\be\label{norma}
\mr{\uD}^\eps(\rho \| \rho) = 0\;.
\ee
Moreover, the DPI of $\mr{\D}^\eps$ is directly inherited from the DPI of $\D$, justifying the term \emph{induced} divergence. To summarize, for any quantum relative entropy $\D$:
\bmyl
$\mr{\uD}^\eps$  is a normalized quantum divergence.
\emyl
\begin{proof}
To verify the normalization property~\eqref{norma}, consider the case where $\sigma = \rho$. In this scenario, we compute:
\ba
\D\left(\rho \| \rho + t\sigma\right) &= \D\left(\rho \| (1+t)\rho\right) \\
\GG{\eqref{scale}} &= \D\left(\rho \| \rho\right) - \log(1+t) \\
&= -\log(1+t)\;.
\ea
Thus, for $\sigma = \rho$, the condition $\D\left(\rho \| \rho + t\sigma\right) \geq \log(1-\eps)$ simplifies to $(1+t)^{-1} \geq 1-\eps$, which directly leads to~\eqref{norma}.

Next, to prove the DPI, let $\rho \in \md(A)$, $\sigma \in \pos(A)$, and $\mN \in \cptp(A \to B)$. Using the DPI property of $\D$, we observe:
\ba
&\sup_{t\in\mbb{R}_+}\big\{t\;:\;\D\left(\mN(\rho)\big\|\mN(\rho)+t\mN(\sigma)\right)\geq {\log}(1-\eps)\big\}\\
&\leq \sup_{t\in\mbb{R}_+}\big\{t\;:\;\D\left(\rho\|\rho+t\sigma\right)\geq {\log}(1-\eps)\big\}\;,
\ea
which implies that $\mr{\D}^\eps$ also satisfies the DPI. This completes the proof.
\end{proof}

Let $\D$ be a quantum relative entropy continuous in its second argument. Then, for all $\rho\in\md(A)$, $\sigma\in\pos(A)$, and $\eps\in(0,1)$ we have:
\bmyl
\ba\label{two}
&\textbf{1.}\quad\quad\mr{\uD}^\eps(\rho\|\sigma)\leq\D(\rho\|\sigma)+\log\left(\frac1\eps\right)\;.\\
&\textbf{2.}\quad\quad\lim_{\eps\to 1^-}\mr{\uD}^\eps(\rho\|\sigma)=\D(\rho\|\sigma)\;.
\ea
\emyl
\begin{proof}
Due to the scaling property~\eqref{scale}, renaming $\lambda+\log((1-\eps)/\eps)$ as $\lambda$, we can express the normalized induced divergence, somewhat more compactly as:
\be\label{compact}
\mr{\uD}^\eps(\rho\|\sigma)=\sup_{\lambda\in\mbb{R}}\left\{\lambda\;:\;\D\left(\rho\big\|(1-\eps)\rho+\eps 2^\lambda\sigma\right)\geq 0\right\}.
\ee
Due to the L\"owner monotonicity of $\D$ (see~\eqref{lm}) we have
\ba
\D\left(\rho\big\|(1-\eps)\rho+\eps 2^\lambda\sigma\right)&\leq\D\left(\rho\big\|\eps 2^\lambda\sigma\right)\\
\GG{\eqref{scale}}&=\D\left(\rho\big\|\sigma\right)-\log(\eps)-\lambda\;.
\ea
Utilizing this inequality in~\eqref{compact} results with
\ba
\mr{\uD}^\eps(\rho\|\sigma)&\leq\sup_{\lambda\in\mbb{R}}\left\{\lambda\;:\;\D\left(\rho\big\|\sigma\right)-\log(\eps)-\lambda\geq 0\right\}\\
&=D(\rho\|\sigma)-\log(\eps)\;.
\ea
This proves the first relation in~\eqref{two}. To prove the second relation, we
combine~\eqref{compact} with the assumption that $\D$ is continuous in its second argument, to get
\ba
\lim_{\eps\to 1^-}\mr{\uD}^\eps(\rho\|\sigma)&=\sup_{\lambda\in\mbb{R}}\left\{\lambda\;:\;\D\left(\rho\|2^\lambda\sigma\right)\geq 0\right\}\\
\GG{\eqref{scale}} &=\D(\rho\|\sigma)\;.
\ea
This completes the proof.
\end{proof}

From the definition of the induced divergence, it follows that if two quantum relative entropies, $\D_1$ and $\D_2$, satisfy the inequality $\D_1(\rho\|\sigma) \leq \D_2(\rho\|\sigma)$ for all $\rho \in \md(A)$ and $\sigma \in \pos(A)$, then the same inequality holds for their corresponding induced divergences. Specifically, for every $\eps \in (0,1)$, $\rho \in \md(A)$, and $\sigma \in \pos(A)$, we have
\be\label{d1d2}
\mr{\D}_1^\eps(\rho\|\sigma)\leq \mr{\D}_2^\eps(\rho\|\sigma)\;.
\ee

Every normalized quantum divergence is non-negative when evaluated on density matrices (see, e.g.,~\cite{GT2020}). Consequently, the lemma implies that $\mr{\uD}^\eps(\rho \| \sigma) \geq 0$ for all $\rho, \sigma \in \md(A)$. However, if $\sigma$ is not a density matrix, $\mr{\uD}^\eps(\rho \| \sigma)$, like any normalized quantum divergence, can take negative values.

The next lemma demonstrate that there exists quantum divergences that are self-induced, meaning $\mr{\uD}^\eps=\D$ for every $\eps\in(0,1)$. Specifically, for every $\eps\in(0,1)$, $\rho\in\md(A)$ and $\sigma\in\pos(A)$, we have:
\bmyl\label{lem9}
\ba
\mr{\underline{D}}^\eps_{\min}(\rho\|\sigma) &= D_{\min}(\rho\|\sigma)\quad\text{and}\\
\mr{\underline{D}}^\eps_{\max}(\rho\|\sigma) &= D_{\max}(\rho\|\sigma)\;.
\ea
\emyl

\begin{proof}
By definition, the induced divergence $\mr{D}_{\min}^\eps$ is given by:
\be\label{val}
\mr{D}_{\min}^\eps(\rho\|\sigma) \eqdef  \log (t)\;,
\ee
where $t$ is the largest value satisfying the condition:
\be\label{h}
D_{\min}\left(\rho\|\rho + t\sigma\right) \geq \log(1-\eps)\;.
\ee
To proceed, denote by $\Pi_\rho$ the projection onto the support of $\rho$. Then,
\ba
D_{\min}\left(\rho\|\rho+t\sigma\right)&=-\log\tr[(\rho+t\sigma)\Pi_\rho]\\
&=-\log\left(1+t\tr[\sigma\Pi_\rho]\right)\\
&=-\log\left(1+t2^{-D_{\min}(\rho\|\sigma)}\right)\;.
\ea
Thus, the condition~\eqref{h} is equivalent to
\be
1 + t  2^{-D_{\min}(\rho\|\sigma)} \leq \frac{1}{1-\eps}\;,
\ee
which can be rewritten as
\be\label{lar}
\log (t) \leq D_{\min}(\rho\|\sigma) + \log\left(\frac{\eps}{1-\eps}\right)\;.
\ee
The largest value of $t$ satisfying~\eqref{lar} occurs when equality holds. Substituting this value of $t$ into~\eqref{val}, we find  
\be
\mr{\underline{D}}^\eps_{\min}(\rho\|\sigma)=D_{\min}(\rho\|\sigma)\;.
\ee

Next, consider the induced divergence of $D_{\max}$. From its definition, the induced divergence $\mr{D}_{\max}^\eps$ is given by:
\be\label{val2}
\mr{D}_{\max}^\eps(\rho\|\sigma) \eqdef  \log (t)\;,
\ee
where $t$ is the largest value satisfying the condition:
\be\label{h}
D_{\max}\left(\rho\|\rho + t\sigma\right) \geq \log(1-\eps)\;.
\ee
To proceed, we compute
\ba
2^{D_{\max}\left(\rho\|\rho + t\sigma\right)}&=\min\big\{s\in\mbb{R}_+\;:\;s(\rho+t\sigma)\geq \rho\big\}\\
&= \min\left\{s\in\mbb{R}_+\;:\;\frac{st}{1-s}\sigma\geq \rho\right\}\\
\Gg{r\eqdef\frac{st}{1-s}}&=\min\left\{\frac{r}{r+t}\;:\;r\sigma\geq \rho\right\}\\
&=\frac{2^{D_{\max}(\rho\|\sigma)}}{2^{D_{\max}(\rho\|\sigma)}+t}
\ea
 Substituting this into~\eqref{h} yields
 \be\label{lar2}
\log (t) \leq D_{\max}(\rho\|\sigma) + \log\left(\frac{\eps}{1-\eps}\right)\;.
\ee
The largest value of $t$ satisfying~\eqref{lar2} occurs when equality holds. Substituting this value of $t$ into~\eqref{val2}, we find  
\be
\mr{\underline{D}}^\eps_{\max}(\rho\|\sigma)=D_{\max}(\rho\|\sigma)\;.
\ee
This completes the proof.
\end{proof}

In~\cite{GT2020} it was shown that every quantum relative entropy is lower bounded by the min-relative entropy and upper bounded by the max relative entropy. Since the induced divergence is in general not additive under tensor products, we cannot use this result to bound the induced divergence between the min and max relative entropies. Nonetheless, this property still holds.

Let $\eps \in (0,1)$, and let $\mr{\uD}^\eps$ be the normalized induced divergence of a quantum relative entropy $\D$. Then, for every $\rho\in\md(A)$ and $\sigma\in\pos(A)$ we have:
\bmyc\label{ars}
\be
D_{\min}(\rho\|\sigma)\leq \mr{\uD}^\eps(\rho\|\sigma)\leq D_{\max}(\rho\|\sigma)\;.
\ee
\emyc

\begin{proof}
Since every quantum relative entropy $\D$ satisfy~\cite{GT2020,GT2021}
\be
D_{\min}(\rho\|\sigma)\leq \D(\rho\|\sigma)\leq D_{\max}(\rho\|\sigma)\;,
\ee
for every $\rho\in\md(A)$ and $\sigma\in\pos(A)$ we get from~\eqref{d1d2} that 
\be
\mr{\underline{D}}_{\min}^\eps(\rho\|\sigma)\leq \mr{\uD}^\eps(\rho\|\sigma)\leq \mr{\underline{D}}_{\max}^\eps(\rho\|\sigma)\;.
\ee
Combining this with Lemma~\ref{lem9} completes the proof.
\end{proof}

In~\cite{GT2020,GT2021} it was shown that if $\sigma=\sum_{x\in[m]}p_x|x\lr x|$ then every quantum relative entropy $\D$ satisfies
\be\label{x}
\D(|x\lr x|\|\sigma)=-\log (p_x)\;.
\ee
In Eq.~(6.101) of~\cite{Gour2025}, this result was extended as follows: Let $t\in[0,1]$, $\rho,\sigma\in\md(A)$, and $\omega\in\md(B)$, with $A$ and $B$ being arbitrary finite dimensional Hilbert spaces. Then, every quantum relative entropy $\D$ satisfies
\ba\label{xx}
\D\left(\rho^A\oplus\0^B\big\|t\sigma^A\oplus(1-t)\omega^B\right)&=
\D\left(\rho^A\big\|t\sigma^A\right)\\
&=\D\left(\rho^A\big\|\sigma^A\right)-\log(t)\;.
\ea
Observe that this property reduce to the scaling property~\eqref{scale} by taking $\omega=\0^B$, and also extend~\eqref{x} by taking $\rho=\sigma=|x\lr x|$ and renaming $t=p_x$.

Since the induced divergence is in general not additive and therefore not a relative entropy, one cannot expect it to satisfy this property. Yet, in the following lemma we show that it does.

Let $\D$ be a quantum relative entropy, $t,\eps\in[0,1]$, $\rho,\sigma\in\md(A)$, and $\omega\in\md(B)$. Then:
\bmyl
\ba
\mr{\D}^\eps&\left(\rho^A\oplus\0^B\big\|t\sigma^A\oplus(1-t)\omega^B\right)\\
&\;\quad\quad\quad\quad\quad\quad=\mr{\D}^\eps\left(\rho^A\big\|\sigma^A\right)-\log(t)\;.
\ea
\emyl
 
\begin{proof}
By definition,
\be\label{f1}
\mr{\D}^\eps\left(\rho^A\oplus\0^B\big\|t\sigma^A\oplus(1-t)\omega^B\right)=\lambda\;,
\ee
where $\lambda\in\mbb{R}$ is the largest number satisfying
\be
\D\left(\rho^A\oplus\0^B\big\|\left(\rho^A+ 2^\lambda t\sigma^A\right)\oplus 2^\lambda(1-t)\omega^B\right)\geq \log(1-\eps).
\ee
Since $\D$ satisfies~\eqref{xx}, this relation is equivalent to
\be\label{x3}
\D\left(\rho^A\big\|\rho^A+2^\lambda t\sigma^A\right)\geq \log(1-\eps)
\ee
Finally, since $2^{\lambda'}\eqdef 2^\lambda t$ is the largest number satisfying the inequality~\eqref{x3} we get that $\lambda'=\mr{\D}^\eps(\rho\|\sigma)$.
Combining this with~\eqref{f1} we conclude that
\ba
\mr{\D}^\eps\Big(\rho^A\oplus\0^B&\big\|t\sigma^A\oplus(1-t)\omega^B\Big)=\log\left(\frac{2^{\lambda'}}t\right)\\
&=\mr{\D}^\eps\left(\rho^A\big\|\sigma^A\right)-\log(t)
\ea
This completes the proof. 
\end{proof}

\subsection{The Induced R\'enyi Divergence}

In this paper we consider primarily the induced divergence of the sandwiched relative entropy $D_\alpha$, which can be expressed as follows: 
For $\alpha\in(1,\infty]$,
\be\label{sandwichid}
\mr{D}^\eps_\alpha(\rho\|\sigma)\eqdef\sup_{\lambda\in\mbb{R}}\left\{\lambda\;:\;Q_\alpha\left(\rho\|\rho+2^\lambda\sigma\right)\geq (1-\eps)^{\alpha-1}\right\}.
\ee
For $\alpha=1$, 
\be\label{idshan}
\mr{D}^\eps(\rho\|\sigma)\eqdef\sup_{\lambda\in\mbb{R}}\left\{\lambda\;:\;D\left(\rho\|\rho+2^\lambda\sigma\right)\geq \log(1-\eps)\right\}.
\ee
For $\alpha\in[0,1)$,
\be\label{sandwichid2}
\mr{D}^\eps_\alpha(\rho\|\sigma)\eqdef\sup_{\lambda\in\mbb{R}}\left\{\lambda\;:\;Q_\alpha\left(\rho\|\rho+2^\lambda\sigma\right)\leq (1-\eps)^{\alpha-1}\right\}.
\ee

Since $D_\alpha$ is monotonically increasing with $\alpha$, $\mr{D}_\alpha^\eps$ is also monotonically increasing in $\alpha$; {\it cf.}~\eqref{d1d2}. 
Additionally, observe that if $\supp(\rho)\subseteq\supp(\sigma)$, then $\mr{D}_\alpha^\eps(\rho\|\sigma)<\infty$. In this case, there always exists a \emph{finite} optimal $t>0$ such that $Q_\alpha(\rho\|\rho+t\sigma)=(1-\eps)^{\alpha-1}$. This property ensures that the induced divergence remains well-behaved under reasonable assumptions on the support of the input states.

The following lemma highlights the relationship between the hypothesis testing divergence  and the induced divergence of the sandwiched R\'enyi relative entropy of order $\alpha \in [0,2]$. 
Let $\rho,\sigma\in\md(A)$, $\eps\in(0,1)$, and $\alpha\in[0,2]$. Then:
\bmyl\label{bounds}
\be\label{aside2}
\mr{D}_\alpha^\eps(\rho\|\sigma)\leq D_H^\eps(\rho\|\sigma)+\log(\eps)\;.
\ee
\emyl
\noindent\textbf{Remark.}
Since the hypothesis testing divergence is not normalized, specifically satisfying $D_H^\eps(\rho\|\rho) = -\log(1-\eps)$,  it is natural to compare the normalized induced R\'enyi divergence with a normalized version of the hypothesis testing divergence. The latter is defined for all $\rho \in \md(A)$, $\sigma \in \pos(A)$, and $\eps \in [0,1)$ as:
\be\label{aside3}
\underline{D}_H^\eps(\rho\|\sigma) \eqdef D_H^\eps(\rho\|\sigma) + \log(1-\eps)\;.
\ee
Thus, in terms of normalized divergences, the inequality in~\eqref{aside2} takes the form:
\be
\mr{\underline{D}}_\alpha^\eps(\rho\|\sigma)\leq \underline{D}_H^\eps(\rho\|\sigma)\;.
\ee

\begin{proof}
Since $\mr{D}_\alpha^\eps$ is increasing in $\alpha$, it is sufficient to prove it for $\alpha=2$.
For every $t\in\mbb{R}_+$ we denote by
\be
\Lambda_t\eqdef(\rho+t\sigma)^{-\frac12}\rho(\rho+t\sigma)^{-\frac12}.
\ee
With this notation 
$Q_2(\rho\|\rho+t\sigma)$ can be expressed as $\tr\left[\rho\Lambda_t\right]$ so that
\be\label{35n}
\mr{D}_2^\eps(\rho\|\sigma)=\sup_{t\in\mbb{R}_+}
\left\{\log(t)\;:\;\tr\left[\rho\Lambda_t\right]\geq 1-\eps\right\}\;.
\ee
By definition, $0\leq\Lambda_t\leq I$ and $\tr[\Lambda_t(\rho+t\sigma)]=1$.
The latter equality gives
\be
t=\frac{1-\tr\left[\rho\Lambda_t\right]}{\tr\left[\Lambda_t\sigma\right]}=\frac{\eps}{\tr\left[\Lambda_t\sigma\right]}\;.
\ee 
Substituting this into~\eqref{35n} gives
\ba
\mr{D}_2^\eps(\rho\|\sigma)&=\log(\eps)-\log\inf_{t\in\mbb{R}_+}
\left\{\tr\left[\Lambda_t\sigma\right]:\tr\left[\rho\Lambda_t\right]= 1-\eps\right\}\\
&\leq \log(\eps)+D_H^\eps(\rho\|\sigma)\;,
\ea
where the inequality follows by replacing the infimum over $\Lambda_t$ with an infimum over \emph{all} $\Lambda\in\eff(A)$.  
This completes the proof.
\end{proof}

Since Corollary~\ref{ars} established that the normalized induced divergence is always no smaller than the min-relative entropy, it follows from the relation in~\eqref{aside3} that, for every $\rho \in \md(A)$, $\sigma \in \pos(A)$, and $\alpha\in[0,2]$
\be
\lim_{\eps \to 0^+} \underline{\mr{D}}_\alpha^\eps(\rho\|\sigma) = D_{\min}(\rho\|\sigma)\;.
\ee
This demonstrates that, at least for $\alpha \in [0,2]$, the normalized induced R\'enyi divergence serves as a smoothed version of the min-relative entropy. In the next lemma we find a lower bound for $\mr{D}_\alpha^\eps$ in terms of the hypothesis testing divergence.

Let $\rho,\sigma\in\md(A)$, $\eps\in(0,1)$, $\alpha\in(1,2]$, and $\delta\in[0,\mu)$, where $\mu\eqdef1-(1-\eps)^{\alpha-1}$. Then:
\bmyl\label{bounds11}
\ba\label{aside}
\mr{D}_\alpha^\eps(\rho\|\sigma)&\geq \tD_{\max}^{1-\mu}(\rho\|\sigma)\\
&\geq D_H^\delta(\rho\|\sigma)+{\log}(\mu-\delta)\;.
\ea
\emyl

\begin{proof}
From the definition in~\eqref{sandwichid} and Lemma~\ref{cool0} we get
\ba\label{abo}
\mr{D}_\alpha^\eps(\rho\|\sigma)&\geq\sup_{\lambda\in\mbb{R}}
\left\{\lambda\;:\;\tr\left(\rho-2^\lambda\sigma\right)_+\geq (1-\eps)^{\alpha-1}\right\}\\
\GG{\eqref{iss}}&=\underline{D}_s^{\mu}(\rho\|\sigma)=\tilde{D}_{\max}^{1-\mu}(\rho\|\sigma)\;.
\ea
This proves the first inequality. The second inequality, follows from~\eqref{is}; specifically, we get that
\be
\tilde{D}_{\max}^{1-\mu}(\rho\|\sigma)\geq D_{H}^{\delta}(\rho\|\sigma)+\log(\mu-\delta)\;.
\ee
This completes the proof.
\end{proof}

The second lower bound in~\eqref{aside} cannot be applied to $\alpha=1$ since in this case $\mu=0$. For this case we show that the induced Umegaki relative entropy can be lower bounded by the so called \emph{measured} R\'enyi relative entropy defined in~\eqref{measured}. 

Let $\rho\in\md(A)$, $\sigma\in\pos(A)$, $\alpha,\eps\in(0,1)$ and set
\be
c\eqdef\frac{1}{1-\alpha}\log\left(\log\frac1{(1-\eps)^{1-\alpha}}\right)\;.
\ee 
Then:
\bmyl\label{bounds4}
\be\label{ber}
\mr{D}^\eps(\rho\|\sigma)\geq D_{\alpha}^{\mbb{M}}(\rho\|\sigma)+c\;.
\ee
\emyl

{\it Remark.}
Due to~\eqref{measured2} we can replace~\eqref{ber} with
\be
\mr{D}^\eps(\rho\|\sigma)\geq D_\alpha(\rho\|\sigma)+\log|\spec(\sigma)|+c\;.
\ee
\begin{proof}
Set $\beta\eqdef1-\alpha$ and consider first the classical case in which $\rho\in\md(A)$ and $\sigma\in\pos(A)$ commute, with eigenvalues $\{p_x\}$ and $\{q_x\}$, respectively. Then, 
\ba
D(\rho\|\rho+t\sigma)&=\sum_{x\in\supp(\p)}p_x\big(\log (p_x)-\log(p_x+tq_x)\big)\\
&=-\sum_{x\in\supp(\p)}p_x\log\left(1+t\frac{q_x}{p_x}\right)\\
\Gg{\log(1+r)\leq\frac{r^\beta}{\beta}}&\geq-\frac{t^\beta}{\beta}\sum_{x\in\supp(\p)}p_x^{1-\beta}q_x^\beta\\
&=-\frac{t^\beta}{\beta}Q_\beta(\rho\|\sigma)\;.
\ea
Combining this with the definition in~\eqref{idshan}, we get that the induced divergence is bounded by 
\ba
&\mr{D}^\eps(\rho\|\sigma)\geq\log\sup_{t\in\mbb{R}_+}\left\{t\;:\;-\frac{t^\beta}{\beta}Q_\beta(\rho\|\sigma)\geq {\log}(1-\eps)\right\}\\
&=D_{1-\beta}(\rho\|\sigma)+\frac1\beta{\log\log}\frac1{(1-\eps)^\beta}\;.
\ea
Since $\beta\eqdef1-\alpha$ we obtain~\eqref{ber}. 

To prove the non-commutative case, let $\mE\in\cptp(A\to X)$, where $X$ is a classical system and observe that from the DPI property of $\mr{D}^\eps$ we get
\ba
\mr{D}^\eps(\rho\|\sigma)&\geq\mr{D}^\eps\left(\mE(\rho)\big\|\mE(\sigma)\right)\\
\GG{\substack{\rm commutative\\\rm case}}&\geq D_\alpha\left(\mE(\rho)\big\|\mE(\sigma)\right)+c\;.
\ea
Since this in equality holds for every $\mE\in\cptp(A\to X)$, it also holds for the supremum over all such $\mE$.
This completes the proof.
\end{proof}

A corollary of Lemmas~\ref{bounds}-\ref{bounds4} is that the induced divergence satisfies the asymptotic equipartition property. Specifically, let $\alpha\in[1,2]$, $\eps\in(0,1)$, $\rho\in\md(A)$ and $\sigma\in\pos(A)$. Then,
\bmyc\label{cor3}
\be
\lim_{n\to\infty}\frac1n\mr{D}_\alpha^\eps\left(\rho^{\otimes n}\big\|\sigma^{\otimes n}\right)=D(\rho\|\sigma)\;.
\ee
\emyc

\begin{proof}
The proof follows directly from the quantum Stein's lemma, as stated in~\eqref{qsl}, combined with the lower and upper bounds provided in Lemmas~\ref{bounds}-\ref{bounds4}.
\end{proof}

\section{Enhanced Position-Based Decoding Lemma}

The position-based decoding lemma is a powerful tool in quantum information theory, widely used to analyze quantum communication tasks and resource redistribution protocols. It provides a framework for quantifying how a subsystem of a quantum state can be decoded by leveraging the structure and position of the system within a larger Hilbert space. This concept has been instrumental in deriving tight bounds for one-shot quantum communication tasks such as quantum state redistribution, entanglement distillation, and channel simulation. In this paper, we revisit the position-based decoding lemma and propose improvements that extend its applicability and tighten its bounds. By refining the mathematical framework underlying the lemma, we enhance its effectiveness and demonstrate its potential to yield stronger achievability results in various quantum information processing tasks.

Let $\rho\in\md(RA)$, $\sigma\in\md(A)$, $n\in\mbb{N}$, and let $A^n=(A_1,\ldots,A_n)$ denotes $n$ copies of $A$. Consider a set of quantum states in $\md(RA^n)$, $\{\tau^{RA^n}_x\}_{x \in [n]}$, where for each $x\in[n]$
\be\label{rcyc}
\tau_x^{RA^n}=\rho^{RA_x}\otimes\sigma^{A_1}\otimes\cdots\otimes\sigma^{A_{x-1}}\otimes\sigma^{A_{x+1}}\otimes\cdots\otimes\sigma^{A_n}\;.
\ee
The position-based decoding lemma, as stated in~\cite{AJW2019,ABS+2023}, asserts that for 
\be\label{n}
n\eqdef\left\lceil\eps2^{D_H^\eps(\rho^{RA}\|\rho^R\otimes\sigma^A)}\right\rceil\;,
\ee 
there exists a POVM $\{\Lambda_x^{RA^n}\}_{x \in [n]}$ with the property that for all $x \in [n]$,
\be\label{f6}
\tr\left[\Lambda_x^{RA^n}\tau^{RA^n}_x\right]\geq 1-6\eps\;.
\ee
In the upcoming lemma, we improve this result by replacing the hypothesis testing divergence with the induced divergence and extending the applicability of the lemma to a broader class of states, which we call pairwise index-symmetric. 

A collection of $n$ density matrices, $\{\tau^{RA^n}_x\}_{x \in [n]} \subset \md(RA^n)$, is said to be \emph{pairwise index-symmetric} if there exist $\rho, \sigma \in \md(RA)$ such that for every $x, y \in [n]$,
\bmyd
\be\label{cas}
\tau_{x}^{RA_{y}}=\begin{cases}
\rho^{RA_x} &\text{if }x=y\\
\sigma^{RA_y} &\text{if }x\neq y\;.
\end{cases}
\ee
\emyd
Observe that the states given in~\eqref{rcyc} are pairwise index-symmetric since they satisfy $\tau_{x}^{RA_{x}} = \rho^{RA_x}$ for all $x\in[n]$, and for $x,y\in[n]$ with $y \neq x$, $\tau_{x}^{RA_{y}} = \rho^R \otimes \sigma^{A_y}$. However, pairwise index-symmetric states are not restricted to the specific form given in~\eqref{rcyc}. In fact, pairwise index-symmetric states form a much larger set of states than those that have the form~\eqref{rcyc}. For example, one can extend the states in~\eqref{rcyc} by letting $\sigma^{A^n}$ be a state symmetric under any permutations of the subsystems of $A^n$, and define
\be
\tau_x^{RA^n}=\rho^{RA_x}\otimes\sigma^{A_1\cdots A_{x-1}A_{x+1}\cdots A_n}.
\ee
Observe that while these states are different than~\eqref{rcyc}, they have the same marginals $\tau_{x}^{RA_{x}} = \rho^{RA_x}$ for all $x\in[n]$, and for $x,y\in[n]$ with $y \neq x$, $\tau_{x}^{RA_{y}} = \rho^R \otimes \sigma^{A_y}$.

In the following result (Lemma~\ref{pbdl}) we improve the position-based decoding lemma in three ways:
\ben
\item Extends its applicability to all pairwise index-symmetric states, not just those of the form~\eqref{rcyc}.
\item Eliminates the unnecessary factor of 6 from~\eqref{f6}, thereby improving the probability of correctly decoding the state.
\item Most importantly, reduces the required value of $n$.
\een

Let $\eps\in(0,1)$, $\rho, \sigma \in \md(RA)$, 
\be
n\eqdef\left\lceil 2^{\mr{D}_2^\eps(\rho^{RA}\|\sigma^{RA})}\right\rceil\;,
\ee
and $\{\tau^{RA^n}_x\}_{x \in [n]} \subset \md(RA^n)$ be an pairwise index-symmetric set of states satisfying~\eqref{cas}. Then: 
\bmyl\label{pbdl}
There exists a POVM $\{\Lambda_x^{RA^n}\}_{x\in[n]}$ with the property that for all $x\in[n]$
\be
\tr\left[\Lambda_x^{RA^n}\tau^{RA^n}_x\right]\geq 1-\eps\;.
\ee
\emyl

\begin{proof}
Let $\eta^{RA^n}\eqdef\sum_{x\in[n]}\tau_x^{RA_n}$ and define
\be
\Lambda^{RA^n}_x\eqdef\eta^{-1/2}\tau_x^{RA^n}\eta^{-1/2}\;.
\ee
Then, from the definition of $Q_2$ we get that
\ba
\tr\left[\Lambda^{RA^n}_x\tau_x^{RA^n}\right]&=Q_2\left(\tau_x^{RA^n}\big\|\eta^{RA^n}\right)\\
\GG{DPI}&\geq Q_2\left(\tau_x^{RA_x}\big\|\eta^{RA_x}\right)\\
\GG{\eqref{cas}}&=Q_2\left(\rho^{RA}\big\|\rho^{RA}+(n-1)\sigma^{RA}\right)\\
&\geq 1-\eps\;,\label{subinto}
\ea
where the last inequality follows from the definition of $\mr{D}_2^\eps$ and the fact that 
\ba
n-1&\leq2^{\mr{D}_2^\eps(\rho\|\sigma)}\\
\GG{\eqref{sandwichid}}&=\sup_{t\in\mbb{R}_+}
\left\{t\;:\;Q_2\left(\rho\|\rho+t\sigma\right)\geq 1-\eps\right\}\;,
\ea 
where we used~\eqref{sandwichid} with $\alpha=2$. 
This completes the proof.
\end{proof}

Observe that from Lemma~\ref{bounds} we have that
\ba
n &\eqdef \left\lceil 2^{\mr{D}_2^\eps(\rho^{RA}\|\sigma^{RA})} \right\rceil \\
\GG{\eqref{aside2}} &\leq \left\lceil \eps 2^{D_{H}^\eps(\rho^{RA}\|\sigma^{RA})} \right\rceil\;.
\ea
Thus, this value of $n$ is smaller than the one given in~\eqref{n}.

\section{Classical Communication over a Quantum Channel}

In this section, we apply the results from the previous sections to classical communication over a quantum channel and describe the task within the framework of resource theories~\cite{CG2019,Gour2025}. This perspective, akin to the approach in~\cite{DHW2008}, not only has pedagogical value but also provides a clearer and more unified understanding of certain QIP tasks.

In the resource-theoretic framework, the ability to transmit classical information via a quantum channel corresponds to extracting a noiseless (i.e., identity) classical channel between Alice and Bob. In the classical setting, the identity channel is equivalent to the completely dephasing channel with respect to the classical basis. Thus, the ``golden unit" of this theory is the completely dephasing channel, denoted $\Delta_m \in \cptp(Z_A \to Z_B)$, where $Z_A$ and $Z_B$ are classical systems on Alice's and Bob's sides, respectively, with $|Z_A| = |Z_B| = m$. This channel satisfies the defining property of a golden unit~\cite{CG2019,Gour2025}:
\be
\Delta_m\otimes\Delta_n\cong\Delta_{mn}\quad\quad\forall\;m,n\in\mbb{N}\;.
\ee

The free operations in this resource theory are characterized by \emph{local superchannels}. Specifically, a superchannel $\Theta$ that maps channels in $\cptp(A \to B)$ to channels in $\cptp(A' \to B')$ is local if it can be expressed as:
\be
\Theta[\mN]\eqdef\mF^{B\to B'}\circ\mN^{A\to B}\circ\mE^{A'\to A}\;,
\ee
where $\mE^{A'\to A}$ and $\mF^{B\to B'}$ are local quantum channels applied on Alice's and Bob's sides, respectively. 

Focusing on the task of transmitting classical information over a quantum channel, the goal is to distill $\Delta_m$ (for the largest possible $m$) using only a local superchannel acting on the given channel $\mN^{A \to B}$. Since classical capacity refers to the ability of a quantum channel to transmit classical information, it can be shown that once the result is established for a classical input system $A$, it extends straightforwardly to the case where $A$ is quantum (see, e.g.,~\cite{Wilde2013,Watrous2018}). Thus, without loss of generality, we assume $A$ is classical. Denoting $A$ by $X$, we let $k \eqdef |X|$ represent the size of this classical input to the channel $\mN^{X\to B}$.

\subsubsection{Resource Monotones}\label{sec}

Before computing the optimal rates for simulating or distilling $\Delta_m$, it is useful to identify the functions that quantify resources in this model. Due to the strict limitation to local operations, the only channels in $\cptp(X \to B)$ that Alice and Bob can freely share (i.e., simulate using free operations) are replacement channels of the form
\be
\mF_\omega(|x\lr x|) \eqdef \omega, \quad \forall x \in [k],
\ee
for some fixed $\omega \in \md(B)$. Consequently, a natural resource monotone can be derived from any channel divergence $\D$. Specifically, the function
\be\label{rm} 
\D(\mN\|\mf)\eqdef\min_{\omega \in \md(B)}\D\left( \mN^{X \to B} \big\| \mF^{X \to B}_\omega\right) 
\ee 
quantifies how far the channel $\mN$ is from the set of free channels in $\cptp(X \to B)$.

We focus on the channel extension of the sandwiched Rényi relative entropy, i.e., $\D=D_\alpha$ for any $\alpha\in[0,\infty]$. Since the input system $X$ is classical, we invoke the definition of $\sigma_\p^{XB}$ from~\eqref{sp} and obtain the following expression for the channel extension of $D_\alpha$ (as defined in~\eqref{cd}):
\begin{align}\label{minimax}
D_\alpha\left( \mN \big\| \mF_\omega\right)
&=\min_{\omega \in \md(B)}\sup_{\p \in \prob(k)} D_\alpha\left(\sigma_\p^{XB} \big\| \sigma^X_\p\otimes\omega^B\right)\nonumber\\
&=I_\alpha(X:B)_\mN\;,
\end{align}
where we applied Sion’s minimax theorem. This observation shows that the relative entropy of a resource is precisely the channel mutual information.

A particularly important case is $\alpha=1$, where the Umegaki relative entropy satisfies the triangle equality (see, e.g., Lemma 7.3 of~\cite{Gour2025}). In this case, the minimization over $\omega\in\md(B)$ in~\eqref{minimax} is attained at $\omega^B = \sigma^B_\p$, yielding
\be \label{131}
I(X:B)_\mN=\sup_{\p \in \prob(k)} D\left(\sigma_\p^{XB} \big\| \sigma^{X}_\p \otimes \sigma^B_\p \right) \;. 
\ee

Beyond channel mutual information, many other resource monotones can be constructed. Among them, the induced Rényi mutual information of order 2 plays a central role in our analysis. Similar to~\eqref{131}, it is defined in terms of the induced divergence for every $\eps \in (0,1)$ as:
\be\label{iind}
\mr{I}_2^\eps(X:B)_\mN \eqdef \sup_{\p \in \prob(k)} \mr{D}_2^\eps \left(\sigma_\p^{XB} \big\| \sigma^{X}_\p \otimes \sigma^B_\p \right)\;, 
\ee
where $\sigma^{XB}_\p$ is defined in~\eqref{sp}. It is straightforward to verify that this function is a resource monotone under local superchannels.

\subsubsection{The Conversion Distance}

The conversion distance quantifies how well the channel $\mN$ can simulate the golden unit $\Delta_m$ through local operations. Specifically, the conversion distance from the channel $\mN^{X\to B}$ to the channel $\Delta_m^{Z_A \to Z_B}$ is defined as
\be\label{tdi} T_{\diamond}\left(\mN\xrightarrow{\text{\tiny LO}}\Delta_m\right) = \inf_{\Theta} \frac{1}{2} \left\|\Theta\left[\mN\right] - \Delta_m\right\|_{\diamond}\;, \ee
where the infimum is taken over all free (i.e., local) superchannels $\Theta$, and LO stands for local operations.

Note that the diamond distance on the right-hand side of~\eqref{tdi} is computed between two classical channels: the golden unit $\Delta_m$ and the channel $\mM \eqdef \Theta[\mN]$. Moreover, the channel $\mM$ can be expressed as
\be\label{ezy} \mM^{Z_A\to Z_B} \eqdef \mF^{B\to Z_B} \circ \mN^{X\to B} \circ \mE^{Z_A\to X}\;, \ee
where $\mF \in \cptp(B \to Z_B)$ is a POVM channel, and $\mE \in \cptp(Z_A \to X)$ is a classical channel. Thus, the minimization in~\eqref{tdi} reduces to a minimization over all such channels $\mE$ and $\mF$. 

Since every classical channel $\mE$ can be written as a convex combination of extreme channels, it suffices to restrict $\mE$ in the optimization of~\eqref{tdi} to extreme classical channels. Thus, without loss of generality, we can assume that $\mE$ satisfies $\mE(|z\lr z|) = |x_z\lr x_z|$ for all $z \in [m]$, where $x_z \in [k]$. Here, $x_z$ is a function of $z$ and can be viewed as the encoding of the message $z$ into system $X$. The sequence $x^m \eqdef (x_1, \ldots, x_m)$ is referred to as a \emph{codebook} of size $m$. Since $k \eqdef |X|$, the number of possible codebooks of size $m$ is $k^m$.

Denoting by $\{\Lambda_y\}_{y\in[n]}$ the POVM associated with $\mF^{Z_B\to B}$, and by $\sigma_x^B\eqdef\mN(|x\lr x|)$ for all $x\in[k]$, we obtain that for every $z\in[m]$
\be\label{ez}
\mM^{Z_A\to Z_B}(|z\lr z|^{Z_A})=\sum_{y\in[m]}p_{y|z}|y\lr y|^{Z_B}
\ee
where
\be
p_{y|z}\eqdef \tr\left[\sigma_{x_z}\Lambda_{y}\right]\;.
\ee
Our task  is therefore to choose the encoding $z\mapsto x_z$ and the decoding POVM $\{\Lambda_y\}_{y\in[m]}$ such that $\mM$ is very close to $\Delta_m$. Observe that the diamond distance between these two channels can be expressed as
\ba
\frac12\left\|\mM-\Delta_m\right\|_\diamond=\frac12\max_{z\in[m]}\|\p_z-\e_z\|_1
\ea
where for each $z\in[m]$, $\p_z\in\prob(m)$ is the probability vector whose components are $\{p_{y|z}\}_{y\in[m]}$ and $\{\e_z\}_{z\in[m]}$ is the standard basis of $\mbb{R}^m$. Now, by definition
\be\label{pzz}
\frac12\|\p_z-\e_z\|_1=1-p_{z|z}\;,
\ee
so that
\be\label{126n0}
T_{\diamond}\left(\mN\xrightarrow{\text{\tiny LO}}\Delta_m\right)=\min_{x^m,\Lambda}\max_{z\in[m]}\left(1-\tr\left[\sigma_{x_z}\Lambda_{z}\right]\right)\;,
\ee
where the minimum is over all encoding $z\mapsto x_z$ (i.e., over all $x^m\in[k]^m$) and all POVMs $\{\Lambda_{z}\}_{z\in[m]}$.

The computation of the conversion distance specified in~\eqref{126n0} poses a significant challenge due to the maximization over all $z \in [m]$. This difficulty arises from our choice of the diamond distance as the metric for measuring channel distance. The underlying issue traces back to Shannon, who addressed a similar problem in the classical setting by replacing the worst-case error probability with the \emph{average} error probability. Translating this idea to our context involves substituting the diamond norm with the trace distance between the normalized Choi matrices of the two channels.
Accordingly, the conversion distance based on this metric is defined as
\ba
T_{c}\left(\mN\xrightarrow{\text{\tiny LO}}\Delta_m\right)=\inf_{\Theta} \frac1{2m}\left\|J_{\Theta\left[\mN\right]}-J_{\Delta_m}\right\|_{1}
\ea
where the infimum is over all local superchannels, and the subscript $c$ emphasizes that this conversion distance is Choi-based.

Using $\mM \eqdef \Theta[\mN]$, as defined in~\eqref{ez}, we can rewrite this expression as
\ba\label{129n}
 \frac1{2m}\left\|J_{\mM}-J_{\Delta_m}\right\|_{1}&=\frac1{2m}\sum_{z\in[m]}\|\p_z-\e_z\|_1\\
\GG{\eqref{pzz}}&=\frac1m\sum_{z\in[m]}(1-p_{z|z})\;.
\ea
Thus, for this alternative choice of conversion distance, we find that
\be\label{126n}
T_{c}\left(\mN\xrightarrow{\text{\tiny LO}}\Delta_m\right)=\min\frac1m\sum_{z\in[m]}\left(1-\tr\left[\sigma_{x_z}\Lambda_{z}\right]\right)
\ee
where the minimum is taken over all encodings $z \mapsto x_z$ (equivalently, over all $x^m \in [k]^m$) and all POVMs $\{\Lambda_{z}\}_{z \in [m]}$.
In Appendix~\ref{shannon}, we show Shannon’s argument that $T_c$ is as effective as $T_{\diamond}$, justifying its use as the primary conversion distance metric.

\subsubsection{Distillation Rate}

Let $\eps\in(0,1)$ and $\mN\in\cptp(X\to B)$. The $\eps$-single-shot distillable communication  is defined as:
\be\label{a4}
\distill^\eps(\mN)\eqdef\max_{m\in[m]}\big\{\log (m)\;:\;T_{c}\left(\mN\xrightarrow{\text{\tiny LO}}\Delta_m\right)\leq\eps\big\}\;.
\ee
Recalling the smoothed version of the induced collision mutual information of the channel, $\mr{I}_2^\eps$, defined in~\eqref{iind}, we obtain
\bmyt\label{thmlub}
\be\label{151}
\distill^\eps(\mN)\geq \mr{I}^{\eps}_{2}(X:B)_\mN\;.
\ee
\emyt

The details of the proof can be found in Appendix~\ref{CCA}. Based on the second lower bound provided in Lemma~\ref{bounds11} for the case $\alpha=2$ (note that in this case $\mu=\eps$), it follows from~\eqref{151} that for every $\delta\in[0,\eps)$ and every $\p\in\prob(k)$
\be
\distill^\eps(\mN) \geq
 D_H^{\delta}\left(\sigma^{XB}_\p\|\sigma^{X}_\p\otimes\sigma^{B}_\p\right) + \log\left(\eps-\delta\right)\;.
\ee
This result matches the lower bound established in Eq.~(12) of~\cite{Cheng2023} (after renaming $\eps-\delta$ as $\delta$). Consequently, the lower bound presented in~\eqref{151} is at least as strong as (and typically stronger than) all previously known lower bounds. 

In addition, observe that based on the first lower bound provided in Lemma~\ref{bounds11} we also get that for every $\p\in\prob(k)$ we have
\be\label{15}
\distill^\eps(\mN) \geq \tD_{\max}^{1-\eps}\left(\sigma^{XB}_\p\|\sigma^{X}_\p\otimes\sigma^{B}_\p\right)\;.
\ee
Utilizing the identity $\tD_{\max}^{1-\eps} = \underline{D}_s^\eps$, we observe that this lower bound provides a refinement of the bound established in Theorem~5.24 of~\cite{DL2015}.

\section{Application 2: Quantum State Redistribution}

\begin{figure}[h]\centering    \includegraphics[width=0.4\textwidth]{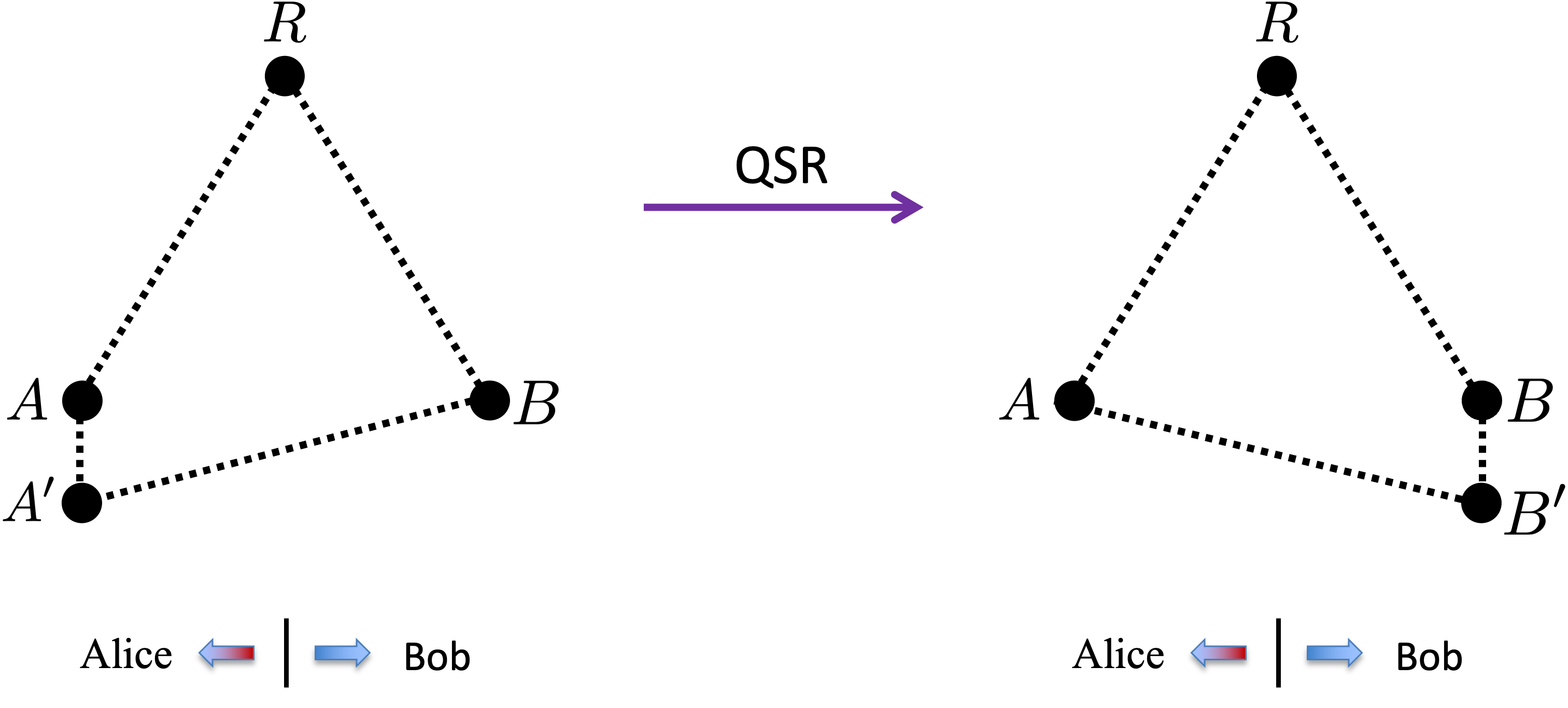}
  \caption{\footnotesize Heuristic description of quantum state redistribution. System $B'$ is a replica of $A'$ on Bob's side.}
  \label{qsr0}
\end{figure}

The second application we consider is quantum state redistribution (QSR). In this setting, a composite system is drawn from the source $\rho^{AA'B}$, where Alice holds two subsystems, $A$ and $A'$, and Bob holds $B$. QSR generalizes quantum state merging to the case where Alice transmits one of her subsystems, $A'$, to Bob while preserving the overall quantum correlations. To analyze this task, we first purify the source $\rho^{AA'B}$ by introducing a reference system $R$, obtaining a four-partite \emph{pure} state $\rho^{RAA'B}$ whose marginal reproduces the original source. The objective of QSR is then to transfer Alice’s subsystem $A'$ of $\rho^{RAA'B}$ to Bob. In resource-theoretic terms, this corresponds to simulating the identity channel $\id^{A'\to B'}$ on $\rho^{RAA'B}$, where $B'$ serves as Bob’s replica of $A'$ (see Fig.~\ref{qsr0}).

In this section, we treat entanglement as a free resource while considering any form of communication -- classical or quantum -- as costly. This setting, known as entanglement-assisted quantum state redistribution (eQSR), permits an arbitrary amount of entanglement in any form, not necessarily as maximally entangled states. Instead of relying on the decoupling theorem, we follow the approach introduced in~\cite{AJW2018} (see also~\cite{ABS+2023} for further developments), which leverages both the convex split lemma~\cite{ADJ2017} and the position-based decoding lemma. By utilizing an equality-based version of the convex split lemma recently introduced in~\cite{Gour2025b}, along with an enhanced position-based decoding lemma (Lemma~\ref{pbdl}) and the induced divergence, we derive tighter bounds than those established in~\cite{AJW2018}.

\begin{figure}[h]\centering    \includegraphics[width=0.4\textwidth]{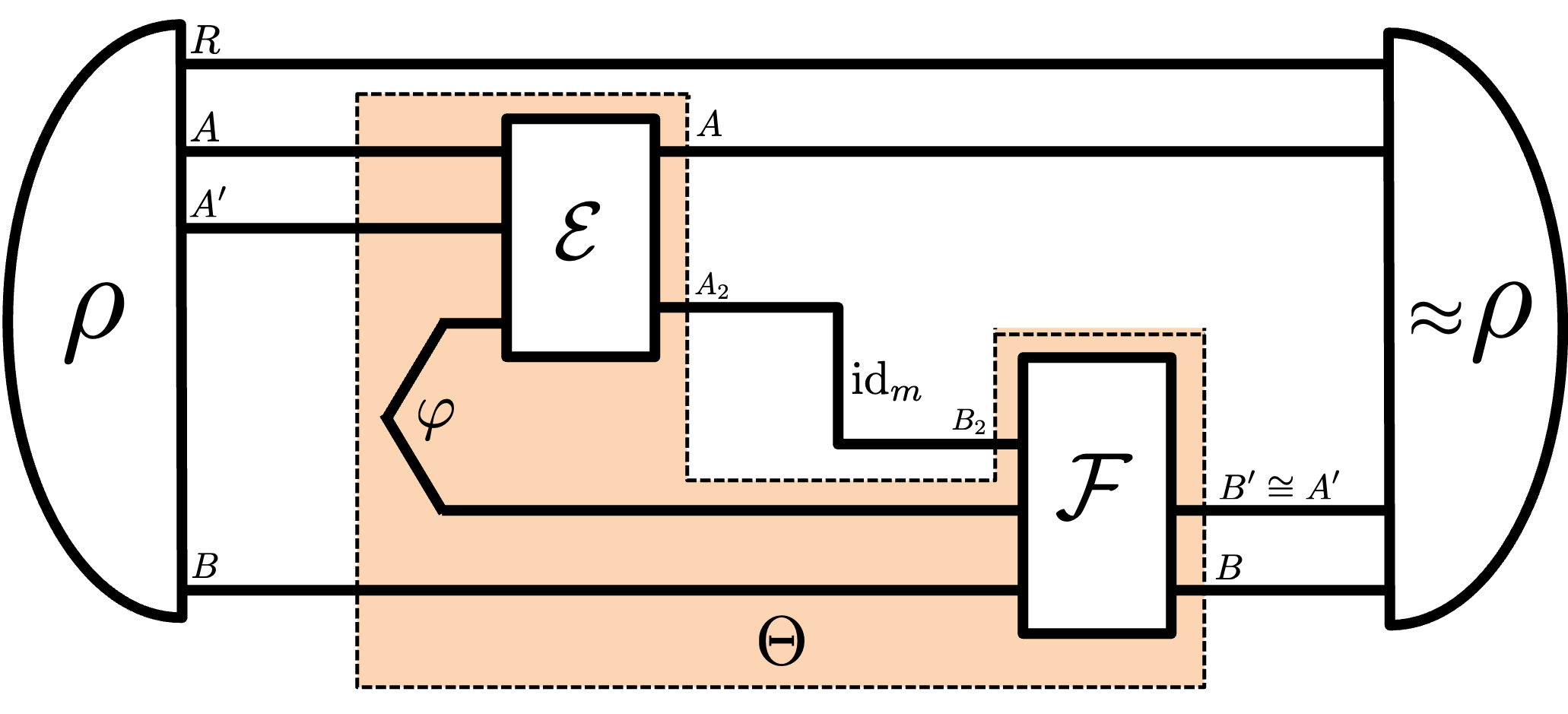}
  \caption{{\footnotesize An $(\eps,q)$-eQSR. The LOSE superchannel $\Theta$ make use of the entangled state $\varphi$.}}
  \label{eqsr}
\end{figure}

In the following definition, we consider a state $\rho\in\pure(RAA'B)$, where $R$ is a reference system, $A$ and $A'$ are held by Alice, and $B$ is held by Bob.
For $\eps\in(0,1)$ and $q\in\mbb{R}+$, we define:
\bmyd
An $(\eps,q)$-eQSR for $\rho^{AA'B}$ is an LOSE superchannel $\Theta$, as in Fig.~\ref{eqsr} with $q=\log (m)$, such that $\mN_\Theta \eqdef \Theta[\id_m]$ satisfies
\be
\frac12\left\|\mN_\Theta^{AA'B\to A B'B}\left(\rho^{RAA'B}\right)-\rho^{RAB'B}\right\|_1\leq\eps\;.
\ee
\emyd

With the above definition of an $(\eps,q)$-eQSR, we can define the $\eps$-error (single-shot) eQSR quantum communication cost as:
\be\label{dfcost}
\cost^{\eps}(\rho^{AA'B})\eqdef\inf\big\{q\;:\;\exists\;(\eps,q)\text{-eQSR for }\rho^{AA'B}\big\}\;.
\ee
Note that we only consider the quantum communication cost as entanglement considered free. 

Let $\bdelta\eqdef(\delta_0,\delta_1)\in(0,1)^2$ (i.e. $\delta_0,\delta_1\in(0,1)$), and $\rho\in\md(RAB)$. We will use the notation:
\be
\mr{I}_2^{\bdelta}(A:R|B)_\rho\eqdef I_{2}^{\delta_0}(RB:A)_\rho-\mr{I}_2^{\delta_1}(B:A)_\rho
\ee
where $I_{2}^{\delta_0}(RB:A)_\rho$ is the smoothed version of the mutual information of order 2, as introduced in~\eqref{i2} and~\eqref{srmi}, and
\be
\mr{I}_2^{\delta_1}(B:A)_\rho\eqdef\min_{\sigma\in\md(A)}\mr{D}_2^{\delta_1}\left(\rho^{AB}\|\sigma^{A}\otimes\rho^B\right)\;.
\ee
The quantity $\mr{I}_2^{\bdelta}(A:R|B)_\rho$ is a type of smoothed, single-shot version, of the quantum conditional mutual information. Indeed, it can be verified that for every $\bdelta\in(0,1)^2$ we have
\be
\lim_{n\to\infty}\frac1n\mr{I}_2^{\bdelta}(A^n:R^n|B^n)_{\rho^{\otimes n}}=I(A:R|B)_\rho\;.
\ee
This property follows from the AEP properties of both the R\'enyi divergence of order 2 and its induced divergence (see Corollary~\ref{cor3}). 

Let $\rho\in\md(AA'B)$ with purification $\rho^{RAA'B}$, and let $\eps\in(0,1)$. Consider $\bdelta\eqdef(\delta_0,\delta_1)\in(0,1)^2$ to be sufficiently small such that
\be\label{deltap}
\delta'\eqdef\eps-\sqrt{2(\delta_0+\delta_1)}-\sqrt{2\delta_0}>0\;.
\ee
Then, the $\eps$-error eQSR cost
of a state $\rho^{AA'B}$ as defined in~\eqref{dfcost} is bounded by
\bmyt\label{thmeqsr}
\be\label{m4}
\cost^{\eps}\left(\rho^{AA'B}\right)\leq \frac12
\mr{I}_2^{\bdelta}(A':R|B)_{\rho}+\log\left(\frac1{\delta'}\right).
\ee
\emyt

The proof of the theorem can be found in Appendix~\ref{Aeqsr}.
The inequality~\eqref{m4} improves upon the results of~\cite{AJW2018} in several ways: First, it replaces $D_{\max}$ with $D_2$. Second, the hypothesis testing divergence is replaced by the induced divergence. Finally, we achieve a separation of the smoothing process, allowing the smoothed conditional mutual information to be expressed simply as the difference between two mutual informations.

\section{Conclusions}

We introduced the induced divergence as a new tool for analyzing fundamental tasks in quantum information theory, demonstrating its utility in refining position-based decoding and improving achievable bounds for quantum communication and state redistribution. This divergence offers a novel perspective on quantum relative entropies, bridging gaps among various quantum divergences.

Our results establish the induced divergence as an effective alternative to the hypothesis testing divergence, particularly in single-shot scenarios. By leveraging its properties, we derived tighter bounds on communication rates and quantum resource costs, leading to improved analytical techniques. In particular, it strengthened the position-based decoding lemma, extending its applicability to a broader class of states and sharpening achievability bounds.

Several open questions remain:
\begin{itemize}
\item Limit as $\eps\to0^+$ for $\alpha > 2$: While we showed that the normalized induced R\'enyi divergence of order $\alpha \in [0,2]$ converges to the min-relative entropy as $\eps\to0^+$, the behavior for $\alpha > 2$ is unknown.
\item Gap between Theorem~\ref{thmlub} and prior bounds: The bound in Theorem~\ref{thmlub} is strictly better than previous results, such as those in~\eqref{15}, but its precise advantage remains unclear.
\item Broader applications: Beyond its application in classical communication over a quantum channels and state redistribution, the induced divergence may have further applications in quantum information science and quantum resource theories, which remain to be explored.
\end{itemize}
Clarifying these questions may further illuminate the significance of the induced divergence. We hope this work inspires further study into its mathematical properties and operational implications in quantum information science.

\begin{acknowledgments}
This research was supported by the Israel Science Foundation under Grant No.\ 1192/24.
\end{acknowledgments}

\bibliographystyle{apsrev4-2}
\bibliography{QRT} 

\onecolumngrid

\appendix

\section*{Appendix}
\setcounter{section}{0} 
\renewcommand{\thesection}{\Alph{section}} 

\section{Shannon's Argument}\label{shannon}

Equation~\eqref{129n} implies that if $\frac1mJ_\mM$ is $\eps$-close to $\frac1mJ_{\Delta_m}$ then at least half of the messages $z\in[m]$ satisfy $1-p_{z|z}\leq2\epsilon$. This means that if there exists a free operation $\Theta$ as above that achieves $\frac1mJ_{\Theta[\mN]}\approx_\eps \frac1mJ_{\Delta_m}$, then Alice can use it to communicate a message of size $\lfloor m/2\rfloor$ in such a way that the maximal error probability $\max_{y\in[m/2]}(1-p_{y|y})\leq2\epsilon$. 
Thus, if $T_{c}\left(\mN\xrightarrow{\small\rm LO}\Delta_m\right)\leq\eps$ then $T_{\diamond}\left(\mN\xrightarrow{\small\rm LO}\Delta_{\lfloor m/2\rfloor}\right)\leq2\eps$, so that we must have
\be\label{tdiam}
T_{\diamond}\left(\mN\to\Delta_{\lfloor m/2\rfloor}\right)\leq 2T_{c}\left(\mN\to\Delta_m\right)\;.
\ee
In other words,, if there exists a communication scheme as above in which Alice communicates $\log(m)$ bits to Bob with average probability of error that does not exceed $\epsilon$, then there is also a communication scheme in which Alice communicates $\log(m/2)=\log(m)-1$ bits with a \emph{maximal} error probability that does not exceeds $\epsilon$. This observation shows that up to one bit of communication cost, the two norms above are essentially equivalent for our purposes. 

Specifically, let $\eps\in(0,1)$ and $\mN\in\cptp(X\to B)$. The $\eps$-single-shot distillable communication  is defined as:
\be
\distill^\eps_{\diamond}(\mN)\eqdef\max_{m\in[m]}\big\{\log m\;:\;T_{\diamond}\left(\mN\to\Delta_m\right)\leq\eps\big\}\;.
\ee
Observe that we are using the diamond distance in the definition of the distillable communication. From~\eqref{tdiam} it follows that with respect to the Choi distance we have
\be\label{1p48}
\distill^\eps_{\diamond}(\mN)\geq\distill^{\eps/2}(\mN)-1\;,
\ee
where
\be\label{1p48}
\distill^\eps(\mN)\eqdef\max_{m\in[m]}\big\{\log m\;:\;T_{c}\left(\mN\to\Delta_{m}\right)\leq\eps\big\}\;.
\ee
We therefore used $\distill^\eps$ to estimate the $\eps$-single-shot distillable communication.

\section{Proof of Theorem~\ref{thmlub}}\label{CCA}

\noindent\textbf{Theorem.}
{\it Let $\mN\in\cptp(X\to B)$ and $\eps\in(0,1)$. Then, 
\be
\distill^\eps(\mN)\geq \mr{I}^{\eps}_{2}(X:B)_\mN+\log\left(\frac\eps{1-\eps}\right)\;,
\ee
where $\mr{I}_2^\eps$ is defined in~\eqref{iind}.}

\begin{proof}
For a given $x^m\in[k]^m$ (here $k\eqdef|X|$) and $z\in[m]$, let
$\eta_{x^m}\eqdef\sum_{z\in[m]}\sigma_{x_z}$ and
\be
\Lambda_z\eqdef \eta^{-1/2}_{x^m}\sigma_{x_z}\eta^{-1/2}_{x^m}\;.
\ee
Observe that $\{\Lambda_z\}_{z\in[m]}$ is indeed a POVM.
By taking this POVM in~\eqref{126n}, we get that the conversion distance is bounded by
\be\label{start}
T_{c}(\mN\xrightarrow{\text{\tiny LO}}\Delta_m)\leq 1-\max_{x^m\in[k]^m}\frac1m\sum_{z\in[m]}Q_2\left(\sigma_{x_z}\big\|\eta_{x^m}\right)\;.
\ee
The next step of the proof involves replacing the maximization over all $x^m\in[k]^m$ with an average. The average is calculated with respect to an i.i.d.\ probability distribution of the form $p_{x^m}\eqdef p_{x_1}\cdots p_{x_m}$, where $\p\eqdef(p_1,\ldots,p_k)\in\prob(k)$ is some probability vector. Implementing this change gives
\be\label{io}
T_{c}(\mN\xrightarrow{\small\rm LO}\Delta_m)\leq1-\max_{\p\in\prob(k)}\frac1m\sum_{z\in[m]}\sum_{x^m\in[k]^m}p_{x^m}Q_2\left(\sigma_{x_z}\big\|\eta_{x^m}\right)\;.
\ee
In order to use the the direct sum property in~\eqref{dsp}, recall that
\be
\sigma^{XB}_\p\eqdef\sum_{x\in[k]}p_x|x\lr x|^X\otimes\sigma_{x}^B\;,
\ee
and denote by $X^m=(X_1,\ldots,X_m)$ a classical system comprizing of $m$-copies of $X$.
For a fixed $\p\in\prob(k)$, we denote for every $z\in[m]$
\be
\tau_z^{X^mB}\eqdef
\sum_{x^m\in[k]^m}p_{x^m}|x^m\lr x^m|^{X^m}\otimes\sigma_{x_z}^B\;,
\ee
and 
\ba
\tau^{X^mB}&\eqdef\sum_{x^m\in[k]^m}p_{x^m}|x^m\lr x^m|^{X^m}\otimes\eta_{x^m}^B\\
&=\sum_{z\in[m]}\tau_z^{X^mB}\;.
\ea
With these notations, due to the direct-sum property of $Q_2$ (see~\eqref{dsp}), we can express the term in~\eqref{io} that appear inside the sum over $z\in[m]$ as
\ba
\sum_{x^m\in[k]^m}p_{x^m}Q_2\left(\sigma_{x_z}\big\|\eta_{x^m}\right)
&=Q_2\left(\tau_z^{X^mB}\big\|\tau^{X^mB}\right)\\
\GG{DPI}&\geq Q_2\left(\tau_z^{X_zB}\big\|\tau^{X_zB}\right)\;,
\ea
where we used the DPI with respect to the trace of all systems $X^m=(X_1,\ldots,X_m)$ except for system $X_z$. By definition, $\tau_z^{X^mB}=\sigma^{X_zB}_\p$ and 
\be
\tau^{X_zB}=\sigma^{X_zB}_\p+(m-1)\sigma^{X_z}_\p\otimes\sigma^B_\p\;,
\ee
so that
\ba
Q_2\left(\tau_z^{X_zB}\big\|\tau^{X_zB}\right)&=Q_2\left(\sigma^{X_zB}_\p\big\|\sigma^{X_zB}_\p+(m-1)\sigma^{X_z}_p\otimes\sigma^B_\p\right)\\
&=Q_2\left(\sigma^{XB}_\p\big\|\sigma^{XB}_\p+(m-1)\sigma^{X}_\p\otimes\sigma^B_\p\right)\;.
\ea
We therefore obtain that for every $m\in\mbb{N}$
\be\label{ub}
T_{c}\left(\mN\xrightarrow{\small\rm LO}\Delta_m\right)\leq1-\max_{\p\in\prob(k)}Q_2\left(\sigma^{XB}_\p\big\|\sigma^{XB}_\p+(m-1)\sigma^{X}_\p\otimes\sigma^B_\p\right)\;.
\ee
Finally, to get the lower bound on the distillable rate we substitute the upper bound in~\eqref{ub} into~\eqref{a4} and obtain
\begin{align}\label{175}
\distill^\eps(\mN)&\geq\max_{\p\in\prob(k)}
\max_{m\in\mbb{N}}\big\{\log (m)\;:\;Q_2\left(\sigma_\p^{XB}\big\|\sigma_\p^{XB}+(m-1)\sigma_\p^{X}\otimes\sigma_\p^B\right)\geq 1-\eps\big\}\nonumber\\
&=\max_{\p\in\prob(k)}\log\left(1+\left\lfloor\frac\eps{1-\eps}2^{\mr{D}_2^{\eps}\left(\sigma_\p^{XB}\|\sigma_\p^{X}\otimes\sigma_\p^{B}\right)}\right\rfloor\right)\nonumber\\
&\geq \max_{\p\in\prob(k)}\mr{D}_2^{\eps}\left(\sigma_\p^{XB}\|\sigma_\p^{X}\otimes\sigma_\p^{B}\right)+\log\left(\frac\eps{1-\eps}\right)\;.
\end{align}
This completes the proof.
\end{proof}

\section{Proof of Theorem~\ref{thmeqsr}}\label{Aeqsr}

\noindent\textbf{Theorem.}
{\it Let $\eps,\delta_0,\delta_1\in(0,1)$ be such that~\eqref{deltap} holds. The $\eps$-error eQSR cost
of a state $\rho^{AA'B}$ as defined in~\eqref{dfcost} is bounded by
\be
\cost^{\eps}\left(\rho^{AA'B}\right)\leq \frac12
\mr{I}_2^{\bdelta}(A':R|B)_{\rho}+\log\left(\frac1{\delta'}\right)\;.
\ee
}
\begin{proof}
Let $B'$ be a replica of $A'$ on Bob's side, and set $\rho^{RBB'}\eqdef\id^{A'\to B'}\big(\rho^{AA'B}\big)$. Moreover, let $\sigma \in \md(B')$ and $\trho \in \mb^{\delta_0}(\rho^{RBB'})$ be such that
\be
I_2^{\delta_0}(RB:B')_\rho=D_{2}\left(\trho^{RBB'}\big\|\trho^{RB}\otimes\sigma^{B'}\right)
\ee
By Uhlmann's theorem, and using the fact that for pure states the trace distance is equal to the purified distance, it follows that there exists a purification $\trho^{RABB'}$ of $\trho^{RBB'}$ satisfying
\be\label{e3}
\frac12\left\|\trho^{RABB'}-\rho^{RABB'}\right\|_1=P\left(\trho^{RBB'},\rho^{RBB'}\right)\leq\sqrt{2\delta_0}\;,
\ee
where we used the fact that the purified distance is no greater than the square root of twice the trace distance.
Our objective is to design an LOSE protocol that simulates the action of the channel $\id^{A' \to B'}$ on the state $\rho^{RAA'B}$. Below, we outline the steps of the protocol, explaining the underlying concepts at each stage. However, instead of applying the protocol to $\rho^{RAA'B}$ we will use the state $\trho^{RAA'B}$ instead. The reason for this choice is to express the upper bound on the cost in terms of smoothed functions. As we will see later, this modification introduces only a small error of order $\sqrt{2\delta_0}$.

Before introducing the steps of the eQSR protocol, we establish key notations and properties. The core idea is to invoke the convex split lemma for the marginal state
$
\trho^{RB}\otimes\big(\sigma^{B'}\big)^{\otimes n}
$,
where $n \in \mbb{N}$ will be determined later in the protocol. We use the notation $B'^n = (B'_1, \dots, B'_n)$ for $n$ copies of $B'$ and define the state
\be
\ttau^{RBB'^n}\eqdef\frac1n\sum_{x\in[n]}\trho^{RBB'_x}\otimes\sigma^{B'_1}\otimes\cdots\otimes\sigma^{B'_{x-1}}\otimes\sigma^{B'_{x+1}}\otimes\cdots\otimes\sigma^{B'_n}\;,
\ee
where $\trho^{RBB'_x}\eqdef\id^{A'\to B'_x}(\trho^{RA'B})$.
From the equality-based version~\cite{Gour2025b} of the convex split lemma~\cite{ADJ2017}, particularly Eq.~(24) of~\cite{Gour2025b}, it follows that
\be\label{pmu1}
P\left(\ttau^{RBB'^n},\trho^{RB}\otimes\big(\sigma^{B'}\big)^{\otimes n}\right)\leq\eps_n\;,
\ee
where $\eps_n\eqdef\sqrt{\frac\mu{\mu+n}}$ and $\mu\eqdef Q_{2}(\trho^{RBB'}\|\trho^{RB}\otimes\sigma^{B'})-1$. However, instead of working with $\trho^{RB}$, we aim to express everything in terms of the purified state $\trho^{RAA'B}$. To achieve this, we use Uhlmann's theorem to rewrite all states in~\eqref{pmu1} with their purifications.

Specifically, let $\phi^{AB'}$ be a purification of $\sigma^{B'}$ in $\pure(AB')$. Since
$\trho^{RAA'B}\otimes\big(\phi^{AB'}\big)^{\otimes n}$ is a purification of $\trho^{RB}\otimes\big(\sigma^{B'}\big)^{\otimes n}$, it is left to introduce the following purification of $\ttau^{RB{B'}^n}$:
\be
\big|\ttau^{R(LA^n)(BB'^n)}\big\ra\eqdef\frac1{\sqrt{n}}\sum_{x\in[n]}|x\ra^L\big|\varphi^{RA^n B'^n B}_x\big\ra\;,
\ee
where for every $x\in[n]$
\be\label{defl0}
\varphi_x^{RA^nC^nB}\eqdef\trho^{RA_x C_x B}\otimes\phi^{A_1C_1}\otimes\cdots\otimes\phi^{A_{x-1}C_{x-1}}
\otimes\phi^{A_{x+1}C_{x+1}}\otimes\cdots\otimes\phi^{A_nC_n}\;.
\ee
By Uhlmann's theorem, there exists an isometry channel $\mV \in \cptp(AA'A^n \to LA^n)$ such that
\ba\label{pmu3}
P\left(\ttau^{R(LA^n)(BB'^n)},\mV\left(\trho^{RAA'B}\otimes\big(\phi^{AB'}\big)^{\otimes n}\right)\right)
&=P\left(\ttau^{RBB'^n},\trho^{RB}\otimes\big(\sigma^{B'}\big)^{\otimes n}\right)\\
\GG{\eqref{pmu1}}&\leq\eps_n\;.
\ea
With these notations and properties established, we are now ready to present the six-step protocol for eQSR. Below, we outline the protocol in a table and provide further details and analysis for each step. Notably, the first two steps follow a similar approach to that used in the proof of quantum state splitting as given in~\cite{ADJ2017,Gour2025b}.

\begin{table}[h]
    \centering
    \renewcommand{\arraystretch}{1.5}
    \begin{tabular}{|p{4.5cm}|p{6cm}|p{7cm}|} 
        \hline
        \textbf{Step} & \textbf{Action} & \textbf{Resulting State} \\
        \hline
        \textbf{1. Initial State} & {\footnotesize Alice and Bob borrow $n$ copies of an (entangled) state $\phi^{A{B'}}$.} &  \centering {\footnotesize$\trho^{RAA'}\otimes(\phi^{A{B'}})^{\otimes n}$} \arraybackslash\\
        \hline
        \textbf{2. Uhlmann Isometry} & {\footnotesize Alice apply the isometry channel $\mV\in\cptp(AA'A^n\to LA^n)$ to her systems.} & \centering {\footnotesize$\mV\left(\trho^{RAA'}\otimes\phi^{\otimes n}\right)\approx_{\eps_n}\ttau^{R(LA^n)(B{B'}^n)}$}  \arraybackslash\\
        \hline
        \multicolumn{3}{|p{17.5cm}|}{\footnotesize \textbf{Details:} The $n$ copies of $\phi^{A{B'}}$ serve as the entanglement resource for the protocol, which is considered free under LOSE. As we will see later, much of this resource acts as a catalyst, since by the end of the protocol, Alice and Bob will still share $n-1$ copies. The relations in~\eqref{pmu1} and~\eqref{pmu3} show that $\mV(\trho^{RAA'B}\otimes\phi^{\otimes n})$ is $\eps_n$-close to $\ttau^{R(LA^n)(B{B'}^n)}$, since purified distance equals trace distance for pure states. Thus, using the local isometry $\mV$ and shared entanglement $\phi^{\otimes n}$, Alice can transform $\rho^{RAA'B}$ into a state $\eps_n$-close to $\ttau^{R(LA^n)(B{B'}^n)}$.} \\
        \hline
        \end{tabular}
    \caption{Six-Step eQSR Protocol: Initial Steps.}
    \label{tab:state_redist}
\end{table}

\FloatBarrier

Suppose now that Alice performs a basis measurement on system $L$ in the basis $\{|x\ra^L\}_{x\in[n]}$ on the state $\ttau^{R(LA^n)(B{B'}^n)}$. Upon obtaining the outcome $x$, the resulting state is $\varphi_x$ as defined in~\eqref{defl0}. From the structure of $\varphi_x$ in~\eqref{defl0}, we observe that if Alice swaps the system (register) $A_x$ with $A_1\cong A$, and Bob swaps the system $B'_x$ with $B'_1\cong {B'}$, the state $\varphi_x^{RA^n{B'}^nB}$ transforms into $\trho^{RA{B'}B}\otimes(\phi^{A{B'}})^{\otimes (n-1)}$. However, Bob can perform this swap only after receiving information about the measurement outcome $x$ from Alice. In other words, using local operations assisted by $\log(n)$ bits of classical communication (or equivalently, due to superdense coding, $\frac{1}{2} \log (n)$ bits of quantum communication), Alice and Bob can transform the state $\ttau^{R(LA^n)(B{B'}^n)}$ into $\trho^{RA{B'}B}\otimes(\phi^{A{B'}})^{\otimes (n-1)}$. Moreover, due to the DPI, it follows from~\eqref{pmu3} that applying the same one-way LOCC to the state $\mV\left(\trho^{RAA'B}\otimes\big(\phi^{A{B'}}\big)^{\otimes n}\right)$ results in a state that is $\eps_n$-close to $\trho^{RA{B'}B}\otimes(\phi^{A{B'}})^{\otimes (n-1)}$. 

However, this protocol is not optimal, as Bob's side contains some information about the positions of the subsystems. Thus, instead of transmitting the full measurement outcome $x\in[n]$, Alice can send only partial information about $x$, thereby reducing the communication cost.

To illustrate this, let $m\in[n]$ be an integer to be determined shortly. Any $x\in[n]$ can be expressed as $x=m\ell+j$, where $\ell\eqdef\left\lceil\frac{x}{m}\right\rceil-1$ and $j\eqdef x-m\ell$. By construction, $j$ belongs to $[m]$.
Now, we embed system $L$ on Alice's side into subsystems $L_0L_1$, such that for all $x\in[n]$, we replace $|x\ra^L$ with $|\ell\ra^{L_0}\big|j\ra^{L_1}$. Under this embedding, the state $\ttau$ transforms into
\be
\big|\ttau^{R(L_0L_1A^n)(B{B'}^n)}\big\ra\eqdef\frac1{\sqrt{n}}\sum_{m\ell+j\in[n]}|\ell\ra^{L_0}|j\ra^{L_1}\big|\varphi^{RA^n {B'}^n B}_{m\ell+j}\big\ra\;.
\ee
With this state at hand, Alice can now measure system $L_0$ and send the outcome $\ell$ to Bob. We will then choose $m$ to be the largest possible value that still allows Bob to decode  (or at least approximate with high accuracy) the remaining information, specifically of system $L_1$ (i.e., $j \in [m]$), using his available systems. The following table outlines the next two steps of the protocol.

\begin{table}[ht]
    \centering
    \renewcommand{\arraystretch}{1.5}
    \begin{tabular}{|p{4.5cm}|p{5cm}|p{8cm}|} 
     \hline
        \textbf{Step} & \textbf{Action} & \textbf{Resulting State} \\
        \hline
        \textbf{3. Alice's Measurement} & {\footnotesize Alice measures the subsystem $L_0$  and communicate the outcome $\ell$ to Bob.}&  {\footnotesize Up to an $\eps_n$-error, the resulting state is \be\label{beg0}\big|\ttau_{\ell}^{R(L_1A^n)(B{B'}^n)}\big\ra\eqdef\frac1{\sqrt{m}}\sum\limits_{j\in[m]}|j\ra^{L_1}\big|\varphi^{RA^n {B'}^n B}_{m\ell+j}\big\ra\ee} \\
        \hline
        \textbf{4. Swap of Subsystems} & {\footnotesize Alice replaces the $m$ systems $\{A_{m\ell+j-1}\}_{j\in[m]}$ with $\{A_j\}_{j\in[m]}$, and Bob replaces the $m$ systems $\{{B'}_{m\ell+j-1}\}_{j\in[m]}$ with $\{{B'}_j\}_{j\in[m]}$.}  & {\footnotesize Up to an $\eps_n$-error, the resulting state is $$\Psi^{R(L_1A^m)(B{B'}^m)}\otimes\left(\phi^{A{B'}}\right)^{\otimes (n-m)}\;,$$ where $\Psi^{R(L_1A^m)(B{B'}^m)}$ is defined in~\eqref{psi}.} \\
        \hline
        \multicolumn{3}{|p{17.5cm}|}{\footnotesize \textbf{Details:} In step 3, Alice measures $L_0$ and sends the result to Bob, requiring only $\frac{1}{2} \log\left( \left\lceil \frac{n}{m} \right\rceil -1 \right)$ quantum communication bits via superdense coding.
Replacing the systems as outlined in step 4 is equivalent to setting $\ell = 0$ in~\eqref{beg0}. For $\ell = 0$, the state factorizes into a tensor product of $(n-m)$ copies of $\phi^{A{B'}}$ and the pure state
\be\label{psi}
\big|\Psi^{R(L_1A^m)(B{B'}^m)}\big\ra\eqdef\frac1{\sqrt{m}}\sum_{j\in[m]}|j\ra^{L_1}\big|\ttau^{RA^m {B'}^m B}_j\big\ra
\ee
where 
\be
\ttau^{RA^m{B'}^mB}_{j}\eqdef\phi^{A_1{B'}_1}\otimes\cdots\otimes\phi^{A_{j-1}{B'}_{j-1}}\otimes\trho^{RA_j {B'}_j B}\otimes\phi^{A_{j+1}{B'}_{j+1}}\otimes\cdots\otimes\phi^{A_m{B'}_m}\;.
\ee
        } \\
        \hline
\end{tabular}
    \caption{Six-Step eQSR Protocol: Middle Steps.}
    \label{tab:state_redist}
\end{table}

In the remaining two steps of the protocol, Alice and Bob leave the $(n-m)$ copies of $\phi^{A{B'}}$ intact, and apply operations on their systems described by the state $\Psi^{R(L_1A^m)(B{B'}^m)}$ in~\eqref{psi}. In the fifth step, Bob utilizes the position-based decoding lemma (Lemma~\ref{pbdl}), to construct a measurement to learn $j$. Specifically,
observe that the marginal of $\ttau^{RA^m{B'}^mB}_{j}$ on Bob side has the form
\be
\ttau^{{B'}^mB}_{j}=\sigma^{{B'}_1}\otimes\cdots\otimes\sigma^{{B'}_{j-1}}\otimes\trho^{{B'}_j B}\otimes\sigma^{{B'}_{j+1}}\otimes\cdots\otimes\sigma^{{B'}_m}
\ee
We also consider the state (without the tilde symbol)
\be
\tau^{{B'}^mB}_{j}\eqdef\sigma^{{B'}_1}\otimes\cdots\otimes\sigma^{{B'}_{j-1}}\otimes\rho^{{B'}_j B}\otimes\sigma^{{B'}_{j+1}}\otimes\cdots\otimes\sigma^{{B'}_m}\;.
\ee
From the position-based decoding lemma (Lemma~\ref{pbdl}), we get that for
\be\label{m}
m=\left\lceil 2^{\mr{D}_2^{\delta_1}(\rho^{B'B}\|\sigma^{B'}\otimes\rho^B)}\right\rceil
\ee
 there exists a POVM $\{\Lambda_j\}_{j\in[m]}$ such that for all $j\in[m]$
\be
\tr\left[\Lambda_j^{{B'}^mB}\tau^{{B'}^mB}_{j}\right]\geq1-\delta_1\;.
\ee
Now, recall that $\trho \in \mb^{\delta_0}(\rho^{RB{B'}})$ so that for every $j\in[m]$
\be
\frac12\left\|\ttau^{{B'}^mB}_{j}-\tau^{{B'}^mB}_{j}\right\|_1=\frac12\left\|\trho^{{B'}_jB}-\rho^{{B'}_jB}\right\|_1\leq\delta_0
\ee
Therefore, the POVM $\{\Lambda_j\}_{j\in[m]}$ also satisfies
\be\label{epsi0}
\lambda_j\eqdef\tr\left[\Lambda_j^{{B'}^mB}\ttau^{{B'}^mB}_{j}\right]\geq1-\delta_0-\delta_1\;.
\ee
That is, our choice of $m$ in~\eqref{m} is defined with respect to the original state $\rho$, but the conclusion on the decoding probability in~\eqref{epsi0} is expressed with respect to $\trho$. The reason for this preference is that the divergence $\mr{D}_2^{\delta_1}(\rho^{{B'}B}\|\sigma^{B'}\otimes\rho^B)$ is already a type of smoothed divergence, and there is no need to ``smooth" it further.
With such a POVM that satisfies~\eqref{epsi0}, 
Bob proceed by applying the isometry $U:{B'}^mB\to {B'}^mBL_2$ given by
\be\label{u}
U^{{B'}^mB\to {B'}^mBL_2}=\sum_{j\in[m]}\sqrt{\Lambda_j}\otimes |j\ra^{L_2}\;,
\ee
to the state in~\eqref{psi}.
        
\begin{table}[h]
    \centering
    \renewcommand{\arraystretch}{1.5}
    \begin{tabular}{|p{4.5cm}|p{5cm}|p{8cm}|} 
    \hline
        \textbf{Step} & \textbf{Action} & \textbf{Resulting State} \\
        \hline       
        \textbf{5. Bob's Measurement: Position-Based Decoding} & {\footnotesize Bob applies the isometry in~\eqref{u}.
 }& {\footnotesize Up to an $(\eps_n+\sqrt{2(\delta_0+\delta_1)})$-error, the resulting state is  $$\varphi^{R(L_1A^m)(B{B'}^mL_2)}\otimes\left(\phi^{A{B'}}\right)^{\otimes (n-m)}\;,$$ where $\varphi^{R(L_1A^m)(B{B'}^mL_2)}$ is defined in~\eqref{varphi}.} \\
        \hline
             \multicolumn{3}{|p{17.5cm}|}{\footnotesize \textbf{Details:}  From~\eqref{psi}, and the definitions of $U$ in~\eqref{u} and $\lambda_j$ in~\eqref{epsi0}, we get that 
\be
U\big|\Psi^{R(L_1A^m)(B{B'}^m)}\big\ra=\frac1m\sum_{j\in[m]}\sqrt{\lambda_j}|j\ra^{L_1}\big|\psi^{RA^m {B'}^m B}_j\big\ra|j\ra^{L_2}\quad\text{with}\quad\big|\psi^{RA^m{B'}^mB}_{j}\big\ra\eqdef\frac1{\sqrt{\lambda_j}}\left(I^{RA^m}\otimes \sqrt{\Lambda_j}\right)\big|\ttau^{RA^m{B'}^mB}_{j}\big\ra\;.
\ee
Combining~\eqref{epsi0} with the gentle measurement lemma, and denoting by $\delta\eqdef\delta_0+\delta_1$, we obtain that the fidelity between $|\ttau^{RA^m{B'}^mB}_{j}\ra$ and $|\psi^{RA^m{B'}^mB}_{j}\ra$ is lower bounded by $\la\psi_j|\ttau_j\ra\geq\sqrt{1-\delta}$.
Thus, the fidelity between the resulting state $U|\Psi^{R(L_1A^m)(B{B'}^m)}\ra$ and the state
\be\label{varphi}
\big|\varphi^{R(L_1A^m)(B{B'}^mL_2)}\big\ra\eqdef\frac1{\sqrt{m}}\sum_{j\in[m]}|j\ra^{L_1}\big|\ttau^{RA^m {B'}^m B}_j\big\ra |j\ra^{L_2}\;,
\ee
is bounded by
\ba
\big\la\varphi^{R(L_1A^m)(B{B'}^mL_2)}\big|U\big|\Psi^{R(L_1A^m)(B{B'}^m)}\big\ra
=\frac1m\sum_{j\in[m]}\sqrt{\lambda_j}\big\la\ttau^{RA^m {B'}^m B}_j\big|\psi^{RA^m {B'}^m B}_j\big\ra\geq1-\delta\;,
\ea
where the inequality follows from \eqref{epsi0} and the fact that $\la\psi_j|\ttau_j\ra\geq\sqrt{1-\delta}$. Translating this to the trace distance, and using the triangle inequality, we conclude that at the end of step 5 the resulting state is $(\eps_n+\sqrt{2(\delta_0+\delta_1)})$-close to the state $\varphi\otimes\phi^{\otimes (n-m)}$.
        } \\
        \hline
\end{tabular}
    \caption{Six-Step eQSR Protocol: Fifth Step.}
    \label{tab:state_redist}
\end{table}

\begin{table}[h]
    \centering
    \renewcommand{\arraystretch}{1.5}
    \begin{tabular}{|p{4.5cm}|p{5cm}|p{8cm}|} 
    \hline
        \textbf{Step} & \textbf{Action} & \textbf{Resulting State} \\
        \hline               
        \textbf{6. Controlled Swap} & {\footnotesize Alice and Bob apply controlled unitary swap operations, conditioned on $|j\ra^{L_1}$ for swapping $A_j$ with $A_1\cong A$, and on $|j\ra^{L_2}$ for swapping ${B'}_j$ with ${B'}_1\cong {B'}$.} & {\footnotesize Up to an $(\eps_n+\sqrt{2(\delta_0+\delta_1)})$-error, the resulting state is $$\trho^{RAB{B'}}\otimes\left(\phi^{A{B'}}\right)^{\otimes n-1}\otimes\Phi^{L_1L_2}_m$$} \\
          \hline
        \multicolumn{3}{|p{17.5cm}|}{\footnotesize \textbf{Details:} Conditioned on her state $|j\ra^{L_1}$, Alice applies a swap between her system $A_j$ and her system $A_1$ which we simply denote as $A$. Similarly,  conditioned on his state $|j\ra^{L_2}$, Bob applies a swap between his system $B'_j$ and his system $B'_1$ denoted as ${B'}$. 
Observe that the action of these local unitary operations transform the state $\varphi\otimes\phi^{\otimes (n-m)}$ to the state $\trho^{RAB{B'}}\otimes\phi^{\otimes n-1}\otimes\Phi^{L_1L_2}_m$.
Therefore, the fidelity of the final state of the protocol with the state above is no less than $1-\delta$. Due to the relation between the trace distance and the fidelity we conclude the output of the protocol is $(\eps_n+\sqrt{2\delta})$-close to the state $\trho^{RAB{B'}}\otimes(\phi^{A{B'}})^{\otimes n-1}\otimes\Phi_m^{L_1L_2}$. Since $\trho^{RAB{B'}}$ is $\sqrt{2\delta_0}$-close to $\rho^{RAB{B'}}$, the final state of the protocol is $(\eps_n+\sqrt{2\delta}+\sqrt{2\delta_0})$-close to the state $\rho^{RAB{B'}}\otimes(\phi^{A{B'}})^{\otimes n-1}\otimes\Phi_m^{L_1L_2}$.}\\
        \hline 
    \end{tabular}
    \caption{Six-Step eQSR Protocol: Final Step.}
    \label{tab:state_redist}
\end{table}

\FloatBarrier

\subsection*{Analysis of the Quantum Communication Cost}
The quantum communication cost of the protocol is bounded by
\be
q=\frac12\log\left( \left\lceil\frac n m\right\rceil-1\right)\leq\frac12\log (n)-\frac12\log (m)\;.
\ee
Recall that $\eps_n\eqdef\sqrt{\frac\mu{\mu+n}}$ and $\mu\eqdef Q_{2}(\trho^{RB{B'}}\|\trho^{RB}\otimes\sigma^{B'})-1$. Fix $\delta'\in(0,1)$, and let $n$ be the smallest integer satisfying $\eps_n\leq\delta'$. By definition, 
\be
n=\left\lceil\mu\left(\frac1{\delta'^2}-1\right)\right\rceil\leq\frac{\mu+1}{\delta'^2}\;.
\ee
Furthermore, from its definition in~\eqref{m} we get that $\log(m)$ is lower bounded by
\ba
\log(m)&\geq \log\left(2^{\mr{D}_2^{\delta_1}\left(\rho^{{B'}B}\|\sigma^{B'}\otimes\rho^B\right)}\right)\\
&=\mr{D}_2^{\delta_1}\left(\rho^{{B'}B}\|\sigma^{B'}\otimes\rho^B\right)
\ea
Thus, denoting by $\eps\eqdef\delta'+\sqrt{2(\delta_0+\delta_1)}+\sqrt{2\delta_0}$ we obtain
\ba
\cost^{\eps}\left(\trho^{AA'B}\right)&\leq\frac12\left\{D_{2}\left(\trho^{RB{B'}}\big\|\trho^{RB}\otimes\sigma^{B'}\right)-\mr{D}_2^{\delta_1}\left(\rho^{{B'}B}\|\rho^B\otimes\sigma^{B'}\right)\right\}
+\log\left(\frac1{\delta'}\right)\\
&\leq \frac12\left\{I_{2}^{\delta_0}(RB:{B'})_\rho-\mr{I}_2^{\delta_1}(B:{B'})_\rho\right\}
+\log\left(\frac1{\delta'}\right)\;.
\ea
This completes the proof.
\end{proof}



\end{document}